\documentclass[a4paper,11pt]{article}
\pdfoutput=1 
\usepackage{jheppub} 
\usepackage[T1]{fontenc} 

\usepackage{epsfig}
\usepackage{graphicx}
\usepackage{dcolumn}
\usepackage{bm}
\usepackage{ltablex,booktabs}
\usepackage{overpic}
\usepackage{subfigure}
\usepackage{float}
\usepackage{color}
\usepackage{amsmath}
\usepackage{mathcomp}
\usepackage{mathrsfs}
\usepackage{multirow}
\usepackage{rotating}
\usepackage{amssymb}
\usepackage{gensymb}
\usepackage{amsmath}
\usepackage{tabularx}
\usepackage{epstopdf}
\usepackage{verbatim}
\title{Amplitude analysis and branching-fraction measurement of \boldmath $D_{s}^{+} \to \pi^{+}\pi^{0}\eta^{\prime}$}
\collaborationImg{\includegraphics[width=0.15\textwidth, angle=90]{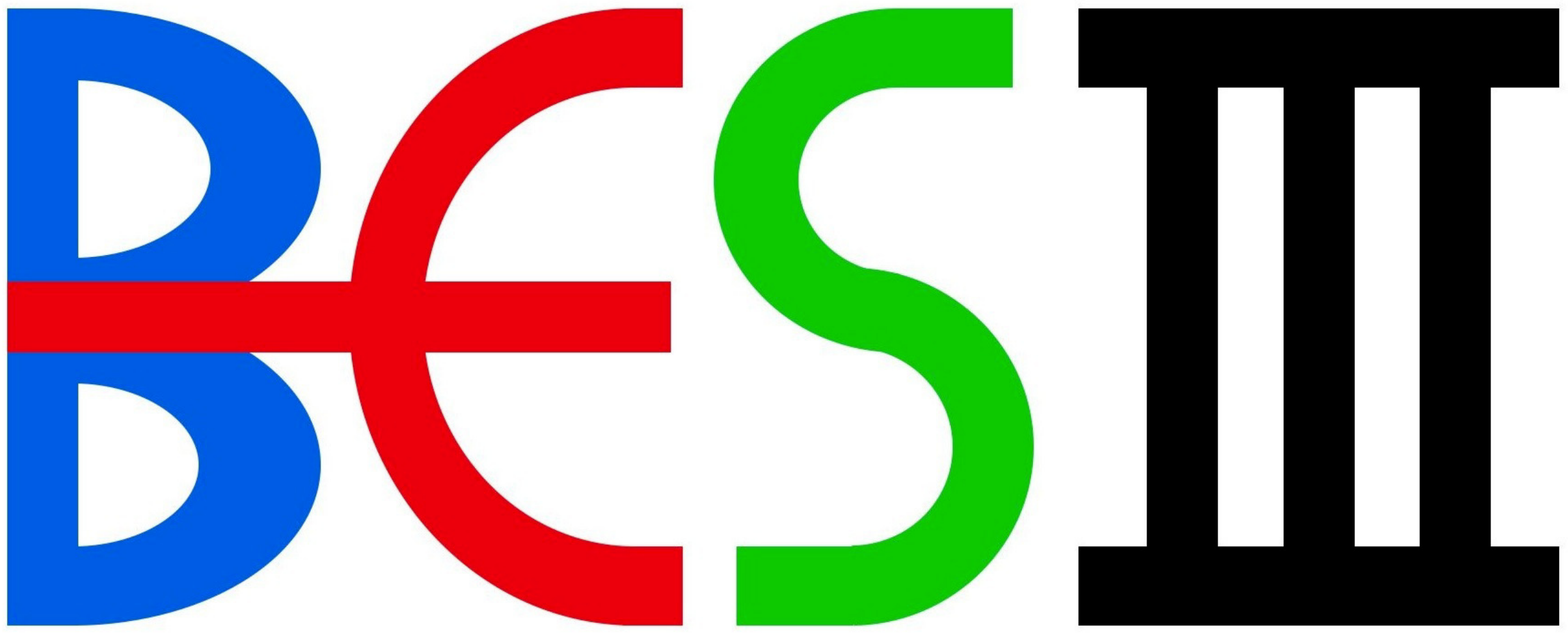}}
\collaboration{BESIII Collaboration}

\date{\today}
\abstract{
  Using data collected with the BESIII detector in $e^+e^-$ collisions at center-of-mass energies between 4.178 and 4.226 GeV and corresponding to 
6.32~fb$^{-1}$ of integrated luminosity, 
we report the amplitude analysis and branching-fraction measurement of the $D^+_s \to \pi^+ \pi^0 \eta^{\prime}$ decay. We find that the dominant intermediate process is $D^+_s \to\rho^+ \eta^{\prime}$ and the significances of other resonant and nonresonant processes are all less than $3\sigma$. The upper limits on the branching fractions of $S$-wave and $P$-wave nonresonant components are set to  0.10\% and 0.74\% at the 90\% confidence level, respectively. In addition, the branching fraction of the $D^+_s \to \pi^+ \pi^0 \eta^{\prime}$ decay is measured to be $(6.15\pm0.25(\rm stat.)\pm0.18(\rm syst.))\%$, which receives significant contribution only from $D_s^+\to \rho^+\eta^{\prime}$ according to the amplitude analysis.
}

\arxivnumber{}
\begin{document}
\maketitle
\flushbottom


\section{Introduction}
Hadronic decays of the  $D_s^\pm$ meson  probe the interplay of short-distance weak-decay matrix elements and long distance QCD interactions. Measurements of the branching fractions (BFs) of these decays provide direct knowledge of the amplitudes and phases in the decay process~\cite{Bhattacharya:2008ke,Cheng:2010ry,Bickert:2020kbn}. In addition, an improved understanding of $D_s^\pm$ decays is particularly valuable for studies of the $B_s^0$ meson, which  mainly decays to final states involving $D_s^\pm$ mesons~\cite{Zyla:2020zbs}. 

\begin{figure}[htbp]
	\centering
	\subfigure{
		\includegraphics[width=4cm]{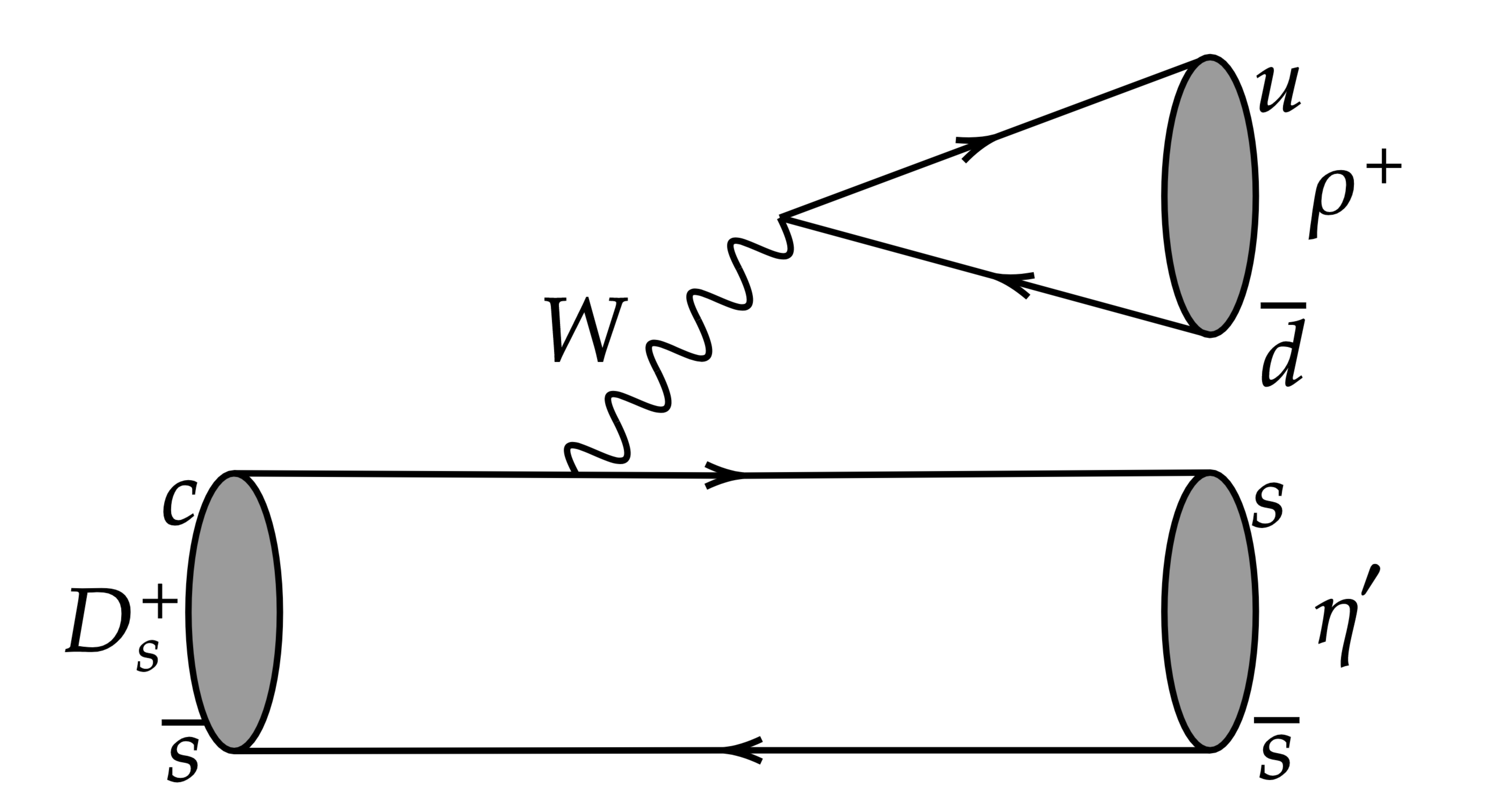}}
	\hspace{0.2cm}
	\subfigure{
		\includegraphics[width=4cm]{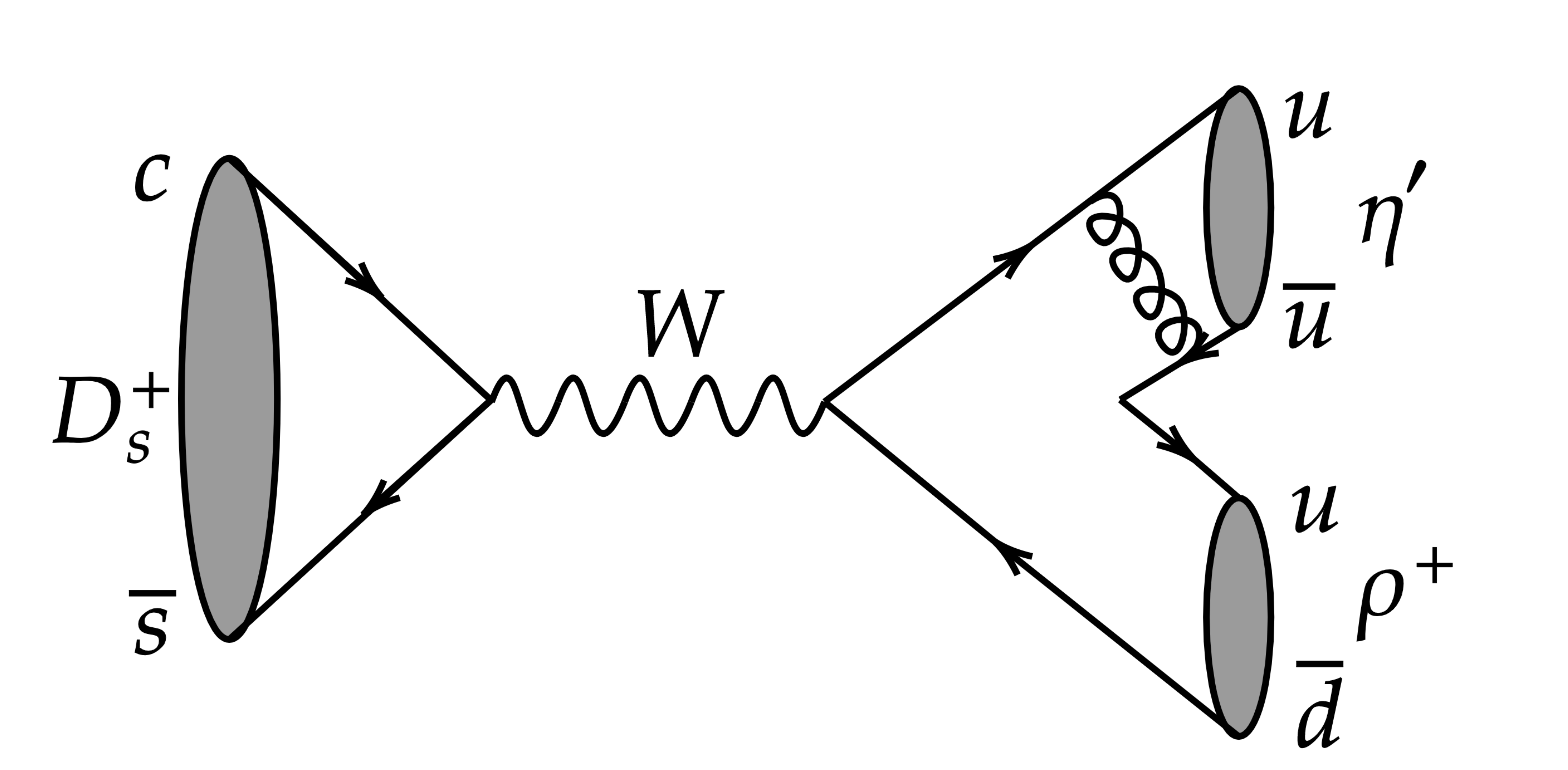}}
	\hspace{0.2cm}
	\subfigure{
		\includegraphics[width=4cm]{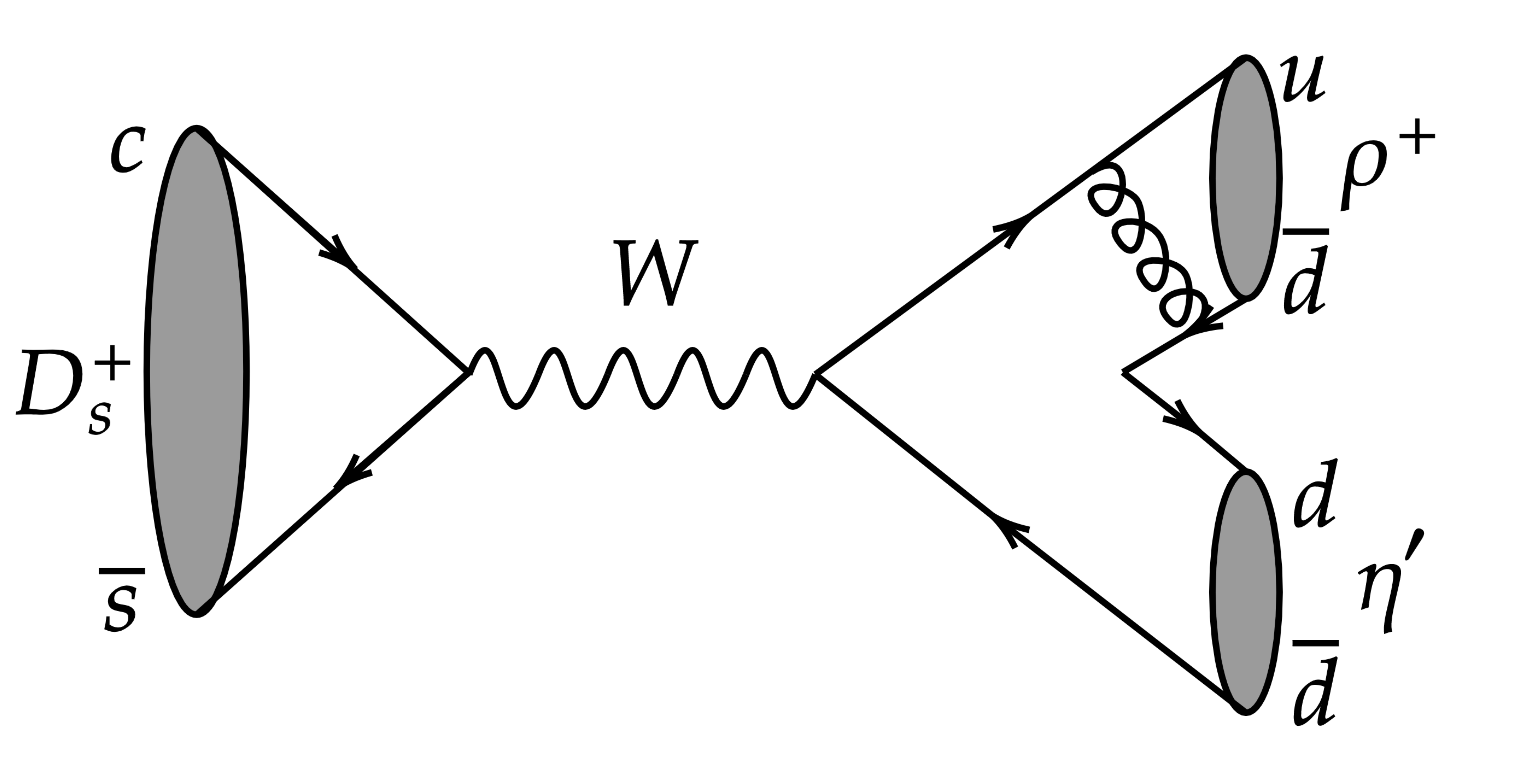}}
	\caption{The $T_P$-diagram (left), $A_V$-diagram (middle) and $A_P$-diagram (right) for $D^+_s\to \rho^+\eta^\prime$. The subscript $P (V )$ implies a pseudoscalar (vector) meson.}
	\label{fig::topo} 
\end{figure}

\begin{figure}[htbp]
	\centering
	\includegraphics[width=4cm]{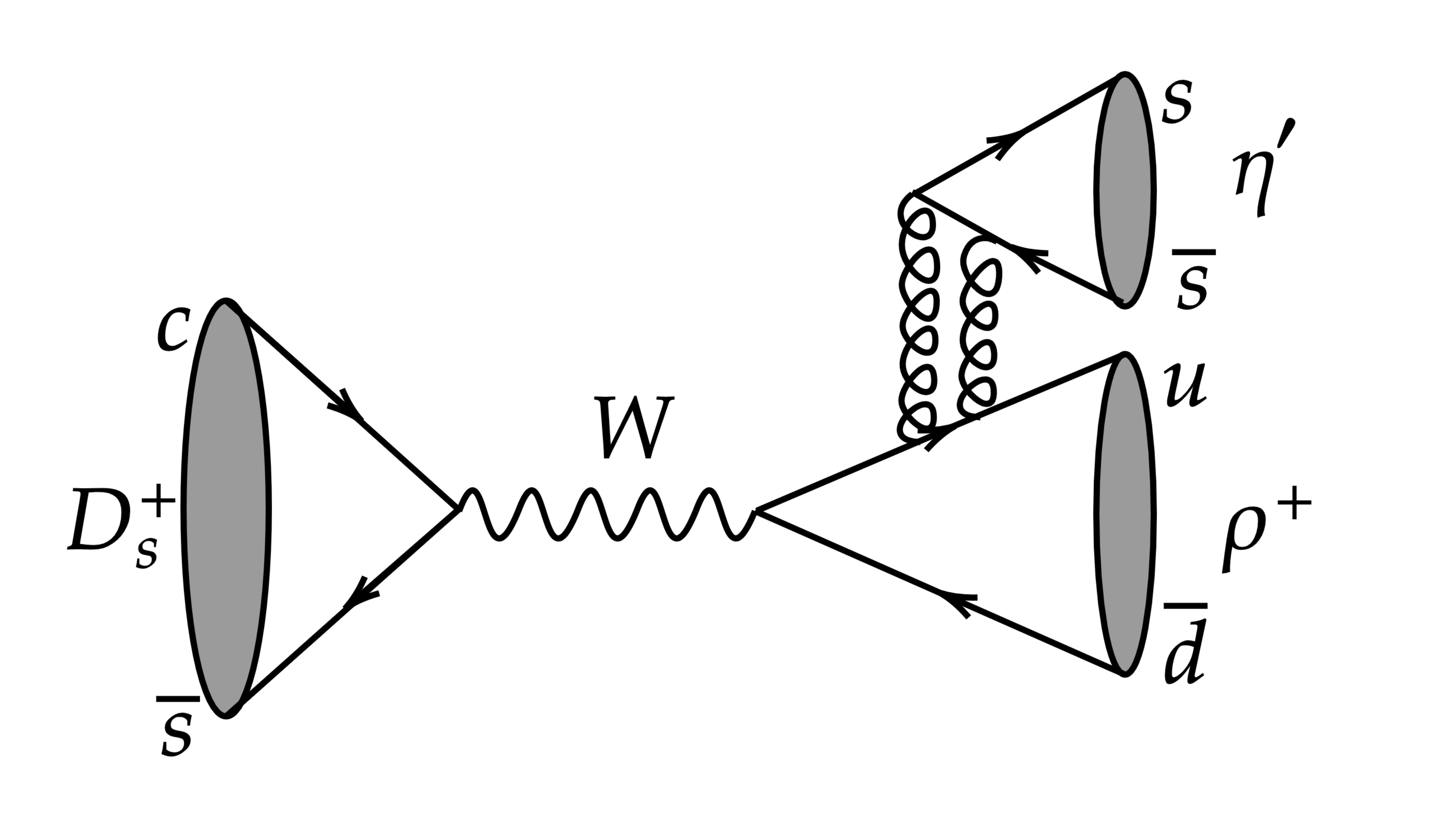}
	\caption{Hairpin-topological diagram for $D^+_s\to \rho^+\eta^\prime$.}
	\label{fig:hair}
\end{figure}

There are two kinds of topological diagrams for $D^+_s\to\rho^+\eta^{\prime}$, including tree ($T$)- and annihilation ($A$)-diagrams, as shown in Fig.~\ref{fig::topo}~\cite{Cheng:2011qh}. Based on reference~\cite{Cheng:2016ejf}, the topological amplitude ($\mathcal{A}$) expressions of $D^+_s\to\rho^+\eta$, $D^+_s\to\rho^+\eta^{\prime}$ and $D^+_s\to\pi^+\omega$ satisfy the sum rule:
\begin{equation} \label{1}
	\frac{1}{\sin\phi}\mathcal{A}(D^+_s\to\pi^+\omega)=\frac{\cos\phi}{\sin\phi}\mathcal{A}(D^+_s\to\rho^+\eta)+\mathcal{A}(D^+_s\to\rho^+\eta^{\prime}).
\end{equation} 
Here, $\phi$ is the mixing angle between $\eta$ and $\eta^{\prime}$ : 
\begin{equation}
	\dbinom{\eta}{\eta^{\prime}}=
\begin{pmatrix}
\cos\phi & -\sin\phi\\
\sin\phi & \cos\phi
\end{pmatrix}
\begin{pmatrix}
\eta_q\\
\eta_s
\end{pmatrix},
\end{equation}
where $\eta_q$ and $\eta_s$ are defined by $\eta_q=\frac{1}{\sqrt{2}}(u\overline{u}+d\overline{d})$ and $\eta_s=s\overline{s}$. Considering the BFs of $D^+_s\to\pi^+\omega$ and $D^+_s\to\rho^+\eta$ and noting a simple triangular inequality in Eq.~(\ref{1}), one obtains the bounds $(2.19\pm0.27)\% < \mathcal{B}(D^+_s\to\rho^+\eta^{\prime}) < (4.51\pm0.38)\%$~\cite{Cheng:2016ejf}. The predictions of the BF of $D^+_s\to\rho^+\eta^{\prime}$ from several theoretical approaches~\cite{Fu-Sheng:2011fji,Qin:2013tje} and the corresponding BFs from experimental measurements are shown in Table~\ref{tab:expBF}. The theoretical predictions for $\mathcal{B}(D^+_s\to\rho^+\eta^{\prime})$ are lower than  the experimental measurement by around $2\sigma$ as shown in Table~\ref{tab:expBF}. A possible way to reconcile the predictions with the measured values would be to take account of the QCD flavor-singlet hairpin contribution shown in Fig.~\ref{fig:hair}~\cite{Cheng:2011qh}. A more precise measurement of the BF of $D^+_s\to\rho^+\eta^{\prime}$ will be very valuable in establishing whether indeed the existing  predictions are incorrect. 
\begin{table}[htpb]
	\caption{$\mathcal{B}(D^+_s\to\rho^+\eta^{\prime})$ from theoretical approaches and previous experimental measurements.}
	\centering
	\label{tab:expBF}
	\begin{tabular}{c|c|ccc}
		\hline
		\hline
		\multicolumn{2}{c}{Decay} &\multicolumn{3}{c}{$\mathcal{B}(\%)$}\\
		\hline
		Theory&$D^+_s\to\rho^+\eta^{\prime}$  &$3.0\pm0.5$~\cite{Fu-Sheng:2011fji} &1.7~\cite{Qin:2013tje}  &1.6~\cite{Qin:2013tje} \\
		\hline
		\multirow{4}{*}{Experiment}&$D^+_s\to\pi^+\pi^0\eta^{\prime}$ & $5.6\pm0.5\pm0.6$  &\multicolumn{2}{|c}{CLEO~\cite{CLEO:2013bae}} \\
		\cline{2-5}
		~&$D^+_s\to\rho^+\eta^{\prime}$ & $5.8\pm1.4\pm0.4$ & \multicolumn{2}{|c}{\multirow{3}{*}{BESIII~\cite{BESIII:2015rrp}}} \\
		\cline{2-3}
		~&$D^+_s\to\pi^+\pi^0\eta^{\prime}$ & $<5.1$ & \multicolumn{2}{|c}{~} \\
		~&(nonresonant) & (90\% confidence level) & \multicolumn{2}{|c}{~}\\
		\hline
		\hline
	\end{tabular}
\end{table}	

 Previously, BESIII reported the BF measurement of $D^+_s\to\rho^+\eta^{\prime}$ performed through the process $e^+e^-\to D_{s}^{+}D_{s}^{-}$, with a 482 pb$^{-1}$ data sample collected at center-of-mass (C.M.) energy $\sqrt s=4.009$ GeV and CLEO measured the BF of $D^+_s\to\pi^+\pi^0\eta^{\prime}$ using 586 pb$^{-1}$ of $e^+e^-$ collisions recorded at C.M. energy $\sqrt s=4.17$ GeV. In this paper, we perform the first amplitude analysis of $D^+_s\to\pi^+\pi^0\eta^{\prime}$ and improve the BF measurement of this decay via the process $e^+e^-\to D_{s}^{*\pm}D_{s}^{\mp}$ by using data samples corresponding to an integrated luminosity of   6.32 fb$^{-1}$  collected by the BESIII detector at C.M. energies $\sqrt{s} = 4.178-4.226$ GeV. Charge-conjugate states are implied throughout this paper.

\section{Detector and data sets}
\label{sec:detector_dataset}
The BESIII detector~\cite{Ablikim:2009aa} records symmetric $e^+e^-$ collisions 
provided by the BEPCII storage ring~\cite{Yu:IPAC2016-TUYA01}, which operates in the C.M. energy range from 2.00 to 4.95~GeV. BESIII has collected large
data samples in this energy region~\cite{Ablikim:2019hff}. The cylindrical core
of the BESIII detector covers 93\% of the full solid angle and consists of a
helium-based multilayer drift chamber~(MDC), a plastic scintillator
time-of-flight system~(TOF), and a CsI(Tl) electromagnetic calorimeter~(EMC),
which are all enclosed in a superconducting solenoidal magnet providing a 1.0~T
magnetic field. The solenoid is supported by an octagonal flux-return yoke with
resistive plate counter muon identification modules interleaved with steel.
The charged-particle momentum resolution at $1~{\rm GeV}/c$ is $0.5\%$, and the specific energy loss (d$E$/d$x$) resolution is $6\%$ for electrons from Bhabha scattering. The EMC
measures photon energies with a resolution of $2.5\%$ ($5\%$) at $1$~GeV in the
barrel (end-cap) region. The time resolution in the TOF barrel region is 68~ps,
while that in the end-cap region is 110~ps. The end-cap TOF
system was upgraded in 2015 using multi-gap resistive plate chamber
technology, providing a time resolution of 60~ps~\cite{etof1, etof2, etof3}.

The data samples used in this analysis are listed in Table~\ref{energe}~\cite{Luminosity}. Since the cross section of $D_{s}^{*\pm}D_{s}^{\mp}$ production in $e^{+}e^{-}$ annihilation is about a
factor of twenty larger than that of $D_{s}^{+}D_{s}^{-}$~\cite{DsStrDs} at C.M. energies $\sqrt{s} = 4.178 - 4.226 $ GeV, and
the $D_{s}^{*\pm}$ meson decays to $\gamma D_{s}^{\pm}$ with a dominant BF of
$(93.5\pm0.7)$\%~\cite{Zyla:2020zbs}, the signal events discussed in this paper are
selected from the process $e^+e^-\to D_{s}^{*\pm}D_{s}^{\mp}\to \gamma D_{s}^{+}D_{s}^{-}$.
 \begin{table}[htb]
 \renewcommand\arraystretch{1.25}
 \centering
 \caption{The integrated luminosities ($\mathcal{L}_{\rm int}$) and the 
   requirements on $M_{\rm rec}$ for various C.M. energies. The
   definition of $M_{\rm rec}$ is given in Eq.~(\ref{eq:mrec}). The first and
   second uncertainties are statistical and systematic, respectively.}
 \begin{tabular}{ccc}
 \hline
 $\sqrt{s}$ (GeV) & $\mathcal{L}_{\rm int}$ (pb$^{-1}$) & $M_{\rm rec}$ (GeV/$c^2$)\\
 \hline
  4.178 &3189.0$\pm$0.2$\pm$31.9&[2.050, 2.180] \\
  4.189 &526.7$\pm$0.1$\pm$2.2&[2.048, 2.190] \\
  4.199 &526.0$\pm$0.1$\pm$2.1&[2.046, 2.200] \\
  4.209 &517.1$\pm$0.1$\pm$1.8&[2.044, 2.210] \\
  4.219 &514.6$\pm$0.1$\pm$1.8&[2.042, 2.220] \\
  4.226 &1056.4$\pm$0.1$\pm7.0$&[2.040, 2.220] \\
  \hline
 \end{tabular}
 \label{energe}
\end{table}

Simulated data samples produced with a {\sc geant4}-based~\cite{GEANT4} Monte
Carlo (MC) package, which includes the geometric description of the BESIII
detector and the detector response, are used to determine detection
efficiencies and to estimate backgrounds. The simulation models the beam energy
spread and initial-state radiation (ISR) in the $e^+e^-$ annihilations with the
generator {\sc kkmc}~\cite{KKMC1, Jadach:1999vf}. The inclusive MC sample includes the
production of open-charm processes, the ISR production of vector
charmonium(-like) states, and the continuum processes incorporated in
{\sc kkmc}. The known decay modes are modelled with
{\sc evtgen}~\cite{EVTGEN1, EVTGEN2} using BFs taken from the Particle Data
Group (PDG)~\cite{Zyla:2020zbs}, and the remaining unknown charmonium decays are modeled with
{\sc lundcharm}~\cite{Hu:2009zzd,Andersson:1999ui,Hu:2001wp, LUNDCHARM1, LUNDCHARM2}. Final-state radiation~(FSR) from
charged final state particles is incorporated using {\sc photos}~\cite{PHOTOS}.

\section{Event selection}
\label{ST-selection}
The data samples were collected just above the $D_s^{*\pm}D_s^{\mp}$ threshold. The tag method~\cite{MARK-III:1985hbd} allows clean signal samples to be selected, providing an opportunity to perform amplitude analyses and to measure the absolute BFs of 
the hadronic $D^+_s$ meson decays. In the tag method, a single-tag~(ST) 
candidate requires only one of the $D_{s}^{\pm}$ mesons to be reconstructed via 
a hadronic decay; a double-tag~(DT) candidate has both $D_s^+D_s^-$ mesons 
reconstructed via hadronic decays. The DT candidates are required to have the 
$D_{s}^{+}$ meson decaying to the signal mode $D_{s}^{+} \to \pi^{+}\pi^{0}\eta^{\prime}$ and the $D_{s}^{-}$ meson decaying to twelve tag modes listed in Table~\ref{tab:tag-cut}. 

\begin{table}[htbp]
 \renewcommand\arraystretch{1.25}
 \centering
 \caption{Requirements on the tagging $D_s^-$ mass ($M_{\rm tag}$) for various tag modes, where the $\eta$
   and $\eta^\prime$ subscripts denote the decay modes used to reconstruct
   these particles.}\label{tab:tag-cut}
     \begin{tabular}{lc}
        \hline
        Tag mode                                     & Mass window (GeV/$c^{2}$) \\
        \hline
        $D_{s}^{-} \to K_{S}^{0}K^{-}$               & [1.948, 1.991]            \\
        $D_{s}^{-} \to K^{+}K^{-}\pi^{-}$            & [1.950, 1.986]            \\
        $D_{s}^{-} \to K_{S}^{0}K^{-}\pi^{0}$        & [1.946, 1.987]            \\
        $D_{s}^{-} \to K_{S}^{0}K^{-}\pi^{-}\pi^{+}$ & [1.958, 1.980]            \\
        $D_{s}^{-} \to K_{S}^{0}K^{+}\pi^{-}\pi^{-}$ & [1.953, 1.983]            \\
        $D_{s}^{-} \to \pi^{-}\eta_{\gamma\gamma}$   & [1.930, 2.000]            \\
        $D_{s}^{-} \to \pi^{-}\eta^{\prime}_{\pi^+\pi^-\eta_{\gamma\gamma}}$  & [1.940, 1.996]            \\
        $D_{s}^{-} \to K^{-}\pi^{+}\pi^{-}$          & [1.953, 1.986]            \\ 
        $D^-_s\to K^-K^+\pi^-\pi^0$                           &[1.947, 1.982]\\
        $D^-_s\to \pi^-\pi^-\pi^+$                            &[1.952, 1.982]\\
        $D^-_s\to \pi^-\eta_{\pi^+\pi^-\pi^0}$                &[1.941, 1.990]\\
        $D^-_s\to \pi^-\eta^{\prime}_{\gamma\rho^0}$                  &[1.939, 1.992]\\
        \hline
      \end{tabular}
\end{table}

Charged tracks detected in the MDC are required to be within a polar angle ($\theta$) range of $|\rm{cos\theta}|<0.93$, where $\theta$ is defined with respect to the $z$-axis which is the symmetry axis of the MDC. For charged tracks not originating from $K_S^0$ decays, the distance of closest approach to the interaction point is required to be less than 10~cm along the beam direction and less than 1~cm in the plane perpendicular to the beam. Particle identification~(PID) for charged tracks combines measurements of the d$E$/d$x$ in the MDC and the flight time in the TOF to form a probability $\mathcal{L}(h)~(h=p,K,\pi)$ for each hadron $h$ hypothesis. Charged kaons and pions are identified by comparing the probability for the two hypotheses, $\mathcal{L}(K)>\mathcal{L}(\pi)$ and $\mathcal{L}(\pi)>\mathcal{L}(K)$, respectively.

The $K^0_S$ candidates are selected by looping over all pairs of tracks with opposite charges, whose distances to the interaction point along the beam direction are within 20 cm. These two tracks are assumed to be pions without PID applied. A primary vertex and a secondary vertex are reconstructed and the decay length between the two vertexes is required to be greater than twice its uncertainty. This requirement is not applied for the $D_s^- \to K^0_S K^-$ decay due to the low combinatorial background. Candidate $K^0_S$ particles are required to have the vertex fit and an
invariant mass of the $\pi^+\pi^-$ pair $(M_{\pi^+\pi^-})$ in the range [0.487, 0.511]~GeV/$c^{2}$. To prevent an event being doubly counted in the $D_s^- \to K^0_S K^-$ and $D_s^-\to K^- \pi^+\pi^-$ selections, the value of $M_{\pi^+\pi^-}$ is required to be outside of the mass range [0.487, 0.511]~GeV/$c^{2}$ for $D_s^-\to K^- \pi^+\pi^-$ decay.

Photon candidates are identified using showers in the EMC. The deposited
energy of each shower must be more than 25~MeV in the barrel
region~($|\rm{cos\theta}|< 0.80$) and more than 50~MeV in the end cap
region~($0.86 <|\rm{cos\theta}|< 0.92$). The opening angle between the position of each
shower in the EMC and the closest extrapolated charged track must be greater
than 10 degrees to exclude showers that originate from charged tracks. The
difference between the EMC time and the event start time is required to be
within [0, 700]~ns to suppress electronic noise and showers unrelated to the
event.

The $\pi^0$ $(\eta)$ candidates are reconstructed through
$\pi^0\to \gamma\gamma$ ($\eta \to \gamma\gamma$) decays, with at least one
photon falling in the barrel region. The invariant mass of the photon pair for $\pi^{0}$ and $\eta$
candidates must be in the ranges $[0.115, 0.150]$~GeV/$c^{2}$ and
$[0.500, 0.570]$~GeV/$c^{2}$, respectively, which are about three times larger than the detector resolution. A kinematic fit that constrains the $\gamma\gamma$
invariant mass to the $\pi^{0}$ or $\eta$ known mass~\cite{Zyla:2020zbs} is performed
to improve the mass resolution. The $\eta$ candidates are also reconstructed through $\eta\to\pi^+\pi^-\pi^0$ and the invariant mass of $\pi^+\pi^-\pi^0$ are required to satisfy the range of $[0.530, 0.560]$~GeV/$c^{2}$. The $\rho^{0}$ candidates are selected via the decay $\rho^{0}\to\pi^{+}\pi^{-}$ with an invariant mass window $[0.620, 0.920]$~GeV/$c^{2}$. The $\eta^{\prime}$ candidates are formed from the
$\pi^{+}\pi^{-}\eta$ and $\gamma \rho^{0}$ combinations with an invariant mass within a range of
$[0.946, 0.970]$~GeV/$c^{2}$.

Twelve tag modes are reconstructed and the corresponding mass windows on the tagging $D_s^-$ mass ($M_{\rm tag}$) are listed in Table~\ref{tab:tag-cut}. The $D^{\pm}_{s}$ candidates with $M_{\rm rec}$ lying within the mass windows listed
in Table~\ref{energe} are retained for further study. The quantity
$M_{\rm rec}$ is the recoil mass of $D^{\pm}_{s}$ and is defined as
\begin{eqnarray}
\begin{aligned}
    M_{\rm rec} = \sqrt{\left(E_{\rm cm} - \sqrt{|\vec{p}_{D_{s}}|^{2}+m_{D_{s}}^{2}}\right)^{2} - |\vec{p}_{D_{s}} | ^{2}} \; , \label{eq:mrec}
\end{aligned}\end{eqnarray}
where $E_{\rm cm}$ is the initial energy of the $e^+e^-$ C.M. system,
$\vec{p}_{D_{s}}$ is the three-momentum of the $D_{s}^{\pm}$ candidate in the
$e^+e^-$ C.M. frame, and $m_{D_{s}}$ is the $D_{s}^{\pm}$ known
mass~\cite{Zyla:2020zbs}.

\section{Amplitude analysis}
\label{Amplitude-Analysis}
\subsection{Further event selection}
\label{AASelection}
The ST $D_{s}^{-}$ mesons are reconstructed using the first eight hadronic decays as shown in Table~\ref{tab:tag-cut} and the following selection criteria are further applied in order to obtain data
samples with high purities for the amplitude analysis. The selection criteria
discussed in this section are not used in the BF measurement since the BF measurement is dominated by statistical uncertainty.

After a tag $D_s^-$ is identified, the signal candidate is selected by requiring one $\eta^{\prime}$ candidate, one track identified as a charged pion and one $\pi^0$ candidate, where $\pi^0$ and $\eta^{\prime}$ candidates are selected by the same requirements described in section~\ref{ST-selection}, but include the decay $\eta^{\prime}\to\pi^+\pi^-\eta$; $\eta\to\gamma\gamma$ only. Then, an nine-constraint (9C) kinematic fit is performed to the process $e^+e^-\to D_{s}^{*\pm}D_{s}^{\mp}\to \gamma D_{s}^{+}D_{s}^{-}$ assuming the $D_s^{-}$ decays to one of the tag modes and the $D_s^{+}$ decays to the signal mode.  Two hypotheses are considered: that the signal $D_s^{+}$ comes from a $D_{s}^{*+}$  meson or the tag $D_s^{-}$ comes from a $D_{s}^{*-}$ meson. The invariant masses of $(\gamma\gamma)_{\pi^{0}}$, $(\gamma\gamma)_{\eta}$, $(\pi^+\pi^0\eta)_{\eta^{\prime}}$, tag $D_{s}^{-}$, and $D_{s}^{*\pm}$ candidates are constrained to the corresponding known masses~\cite{Zyla:2020zbs} and the constraints of four-momentum conservation in the $e^+e^-$ C.M. system are also applied. The $D_s^{*\pm}D_s^{\mp}$ combination with the minimum $\chi_{\rm 9C}^2$ is chosen. In order to ensure that all candidates fall within the phase-space boundary, the constraint of the signal $D_s^{+}$ mass is added to the 9C kinematic fit and the updated four-momenta are used for the amplitude analysis.

To suppress background from fake $\eta$ candidates, we check the invariant-mass distributions of the $\gamma \gamma$ combination ($M_{\rm recombined}$) which can be with one photon from the signal $\eta$ ($\eta_{\rm sig}$) and the other photon  from the signal $\pi^{0}$, $D_{s}^{*}$ or the $\eta/\pi^{0}$ on the tag side. Events with $|M_{\rm recombined}-M_{\pi^{0}}|< 0.015$~GeV/$c^2$ and $|M_{\rm recombined}-M_{\pi^{0}}|< |M_{\rm \eta_{sig}}-M_{\eta}|$ are rejected, where $M_{\rm \eta_{sig}}$ is the invariant mass of the photon pair for $\eta_{sig}$ candidates, while $M_{\pi^{0}}$ and $M_{\eta}$ are the $\pi^0$ and $\eta$ known masses~\cite{Zyla:2020zbs}, respectively. 

\begin{figure*}[!htbp]
  \centering
  \includegraphics[width=0.45\textwidth]{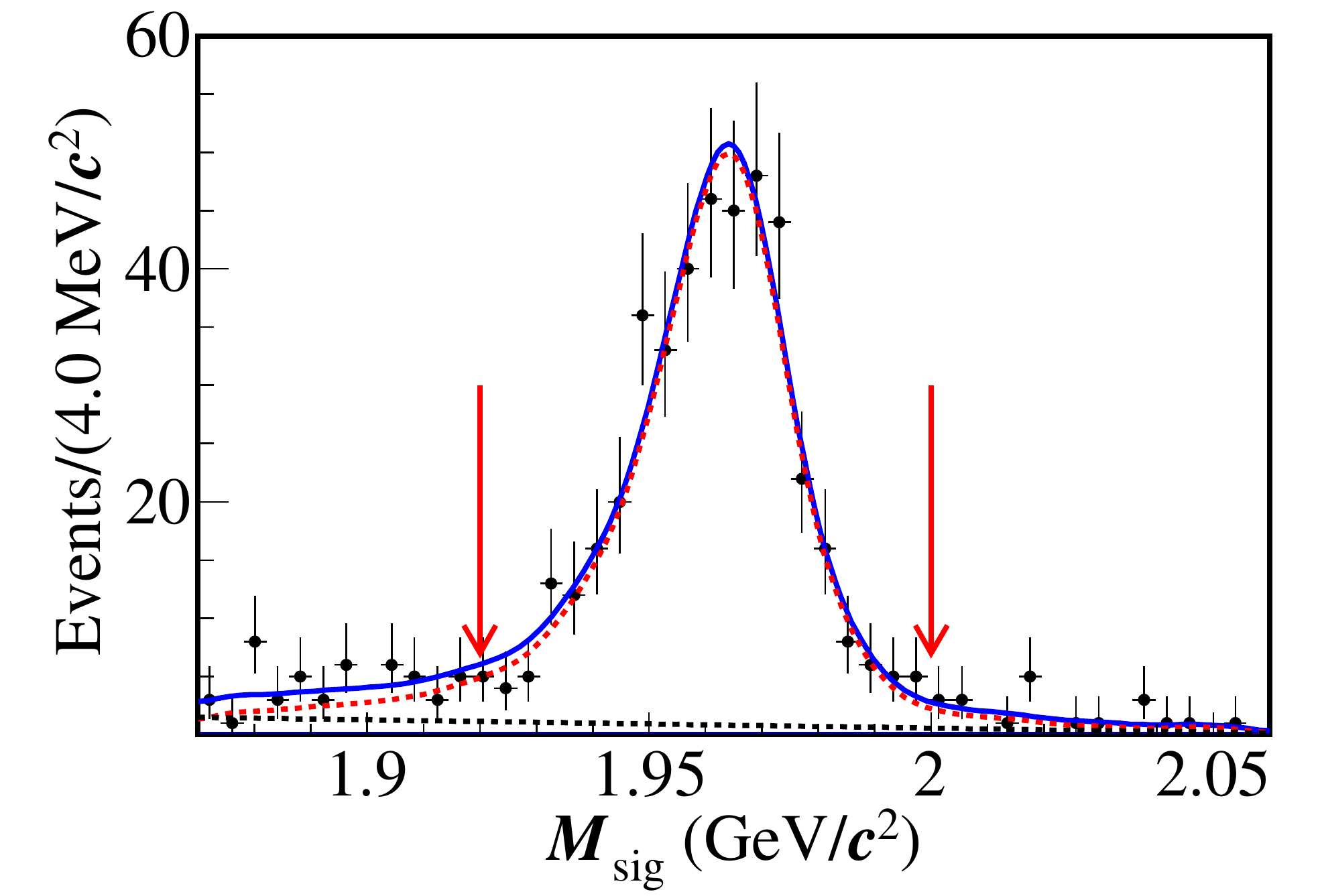}
  \caption{
    Fit to the $M_{\rm sig}$ distribution of the data samples at $\sqrt{s}=$ 4.178-4.226~GeV. This plot is obtained using first eight tag modes in Table~\ref{tab:tag-cut}. The black points with error bars are data. The blue line is the total fit. The red and black dashed lines are the fitted signal and background, respectively. The pair of red arrows indicate the selected signal region.
  } \label{fig:fit_Ds}
\end{figure*}

Figure~\ref{fig:fit_Ds} shows the fits to the invariant-mass distributions of the accepted signal $D_s^{+}$
candidates, $M_{\rm sig}$, for the data samples at $\sqrt{s}=$4.178-4.226~GeV. The signal is
described by a MC-simulated shape convolved with a Gaussian resolution
function, and the background is described by a linear function. Finally, a mass window, $[1.92, 2.00]$~GeV/$c^2$, is applied on the signal $D_s^{+}$ candidates. A total of 411 events are retained for the amplitude analysis with a purity, $w_{\rm sig}$, of $(96.1\pm0.9)\%$.

\subsection{Fit method}
The composition of intermediate resonances in the decay $D_{s}^{+}\to \pi^+\pi^0\eta^{\prime}$
is determined by an unbinned maximum-likelihood fit to data. The likelihood
function is constructed with a probability density function (PDF), which depends
on the momenta of the three daughter particles. The amplitude of the $n^{\rm th}$
intermediate state~($\mathcal{A}_{n}$) is given by
\begin{eqnarray}
\begin{aligned}
  \mathcal{A}_{n} = P_{n}S_{n}F_{n}^{r}F_{n}^{D_{s}}, \label{base-amplitude}
\end{aligned}\end{eqnarray}
where $S_{n}$ and $F_{n}^{r(D_{s})}$ are the spin factor and the Blatt-Weisskopf
barriers of the intermediate state (the $D_{s}^{\pm}$ meson), respectively, and
$P_{n}$ is the propagator of the intermediate resonance.

The nonresonant amplitudes $\mathcal{N}_L$ where $L$ denotes the orbital angular momentum between the $\pi^+ \pi^0$ system  with $L = 0~(S\ \text{wave}), 1~(P\ \text{wave}), 2~(D\ \text{wave})$ are similar to $\mathcal{A}_n$ in Eq.~(\ref{base-amplitude}) but do not contain resonant propagator terms $P_n$: 
\begin{equation}
\mathcal{N}_L = S_nF^r_nF^{D_s}_n,
\end{equation}

The total amplitude $\mathcal{M}$ is then the coherent sum of the amplitudes of intermediate processes, $\mathcal{M} = \sum{\rho_ne^{i\phi_n}\mathcal{A}_n}+\sum{\rho_Le^{i\phi_L}\mathcal{N}_L}$, where the parameters $\rho_n$ and $\phi_n$ are the magnitudes and phases of the $n^{\rm th}$ resonance, while $\rho_L$ and $\phi_L$ correspond to the magnitudes and phases of the nonresonant contribution with angular momentum $L$.

The signal PDF $f_{S}(p_{j})$ is written as 
\begin{eqnarray}\begin{aligned}
  f_{S}(p_{j}) = \frac{\epsilon(p_{j})\left|\mathcal{M}(p_{j})\right|^{2}R_{3}(p_{j})}{\int \epsilon(p_{j})\left|\mathcal{M}(p_{j})\right|^{2}R_{3}(p_{j})\,dp_{j}}\,, \label{signal-PDF}
\end{aligned}\end{eqnarray}
where $\epsilon(p_{j})$ is the detection efficiency parameterized in terms of
the final four-momenta $p_{j}$. The index $j$ refers to the different particles
in the final states, and $R_{3}(p_{j})$ is the standard element of three-body
phase space. The normalization integral is determined by a MC integration,
\begin{eqnarray}\begin{aligned}
  \int \epsilon(p_{j})\left|\mathcal{M}(p_{j})\right|^{2}R_{3}(p_{j})\,dp_{j} \approx
\frac{1}{N_{\rm MC}}\sum_{k}^{N_{\rm MC}} \frac{\left|\mathcal{M}(p_{j}^{k})\right|^{2}}{\left|\mathcal{M}^{g}(p_{j}^{k})\right|^{2}}\,, \label{MC-intergral}
\end{aligned}\end{eqnarray}
where $k$ is the index of the $k^{\rm th}$ event and $N_{\rm MC}$ is the number of the
selected MC events. Here $\mathcal{M}^{g}(p_{j})$ is the PDF used to generate the MC
samples in MC integration. To account for any bias caused by differences in tracking and PID efficiencies, and $\pi^0$ and $\eta$ reconstruction efficiencies between data and MC simulation, each MC event is weighted with a ratio, $\gamma_{\epsilon}(p)$, between the efficiency of data and MC simulation. Then the MC integral becomes
\begin{eqnarray}\begin{aligned}
    &\int \epsilon(p_{j})\left|\mathcal{M}(p_{j})\right|^{2}R_{3}(p_{j})\,dp_{j} \approx
&\frac{1}{N_{\rm MC}} \sum_{k}^{N_{\rm MC}} \frac{\left|\mathcal{M}(p_{j}^{k})\right|^{2}\gamma_{\epsilon}(p_{j}^{k})}{\left|\mathcal{M}^{g}(p_{j}^{k})\right|^{2}}\,.
\label{MC-intergral-corrected}
\end{aligned}\label{eq::gamma}\end{eqnarray}

A signal-background combined PDF is introduced to account for the background in
this analysis. The background PDF is given by
\begin{eqnarray}\begin{aligned}
  f_{B}(p_{j}) = \frac{B(p_{j})R_{3}(p_{j})}{\int B(p_{j})R_{3}(p_{j})\,dp_{j}}\,.\label{bkg-PDF}
\end{aligned}\end{eqnarray}
The background events in the signal region from the inclusive MC sample are
used to model the corresponding background in data. 
This background description is validated by comparing
the $M_{\pi^+\eta^{\prime}}$, $M_{\pi^+\pi^0}$ and
$M_{\pi^0\eta^{\prime}}$ distributions of events outside the $M_{\rm sig}$ signal
region between the data and the inclusive MC samples. The distributions of
background events from the inclusive MC sample within and outside the
$M_{\rm sig}$ signal region are also examined. They are found to be compatible
within statistical uncertainties. The background shape $B(p_{j})$ is a probability density function sampled from a multidimensional histogram by using RooHistPdf implemented in RooFit~\cite{Verkerke}.
This background PDF is then added to the signal PDF incoherently and the
combined PDF is written as
\begin{eqnarray}
\begin{aligned}
  w_{\rm sig}f_{S}(p_{j})&+(1-w_{\rm sig})f_{B}(p_{j})\\
  &=w_{\rm sig}\frac{\epsilon(p_{j})\left|\mathcal{M}(p_{j})\right|^{2}R_{3}(p_{j})}{\int \epsilon(p_{j})\left|\mathcal{M}(p_{j})\right|^{2}R_{3}(p_{j})\,dp_{j}}+(1-w_{\rm sig})\frac{B(p_{j})R_{3}(p_{j})}{\int B(p_{j})R_{3}(p_{j})\,dp_{j}}\,. \label{combined-PDF}
\end{aligned}
\end{eqnarray}
A efficiency-corrected background shape,
$B_{\epsilon}(p_{j})\equiv B(p_{j})/\epsilon(p_{j})$ is introduced in order to
factorize the $\epsilon(p_{j})$ term out from the combined PDF. In this way,
the $\epsilon(p_{j})$ term, which is independent of the fitted variables, is
regarded as a constant and can be dropped during the log-likelihood fit. As a
consequence, the combined PDF becomes
\begin{eqnarray}
\begin{aligned}
  w_{\rm sig}f_{S}(p_{j})+(1-w_{\rm sig})&f_{B}(p_{j})\\
  =\epsilon(p_{j})R_{3}(p_{j})&\left[\frac{w_{\rm sig}\left|\mathcal{M}(p_{j})\right|^{2}}{\int \epsilon(p_{j})\left|\mathcal{M}(p_{j})\right|^{2}R_{3}(p_{j})\,dp_{j}}+\frac{(1-w_{\rm sig})B_{\epsilon}(p_{j})}{\int \epsilon(p_{j})B_{\epsilon}(p_{j})R_{3}(p_{j})\,dp_{j}}
\right]\,. \label{combined-PDF-2}
\end{aligned}
\end{eqnarray}
Next, the integration in the denominator of the background term can also be
handled by the MC integration method in the same way as for the signal only
sample:
\begin{eqnarray}\begin{aligned}
  \int \epsilon(p_{j})B_{\epsilon}(p_{j})R_{3}(p_{j})\,dp_{j} \approx
\frac{1}{N_{\rm MC}}\sum_{k}^{N_{\rm MC}} \frac{B_{\epsilon}(p_{j}^{k})}{\left|\mathcal{M}^{g}(p_{j}^{k})\right|^{2}}\,.
\end{aligned}\end{eqnarray}
The final log-likelihood function is written as 
\begin{eqnarray}\begin{aligned}
  \ln{\mathcal{L}} = \sum_{k}^{N_{D}} \ln\left[w_{\rm sig}f_{S}(p_{j}^{k})+(1-w_{\rm sig})f_{B}(p_{j}^{k})\right]\,,  \label{likelihood3}
\end{aligned}\end{eqnarray}
where $N_{D}$ is the number of candidate events in data.

\subsubsection{Blatt-Weisskopf barrier factors}
For the process $a \to bc$, the Blatt-Weisskopf barrier $F_L(p_j)$~\cite{PhysRevD.104.012016} is
parameterized as a function of the angular momenta $L$ and the momenta $q$ of
the daughter $b$ or $c$ in the rest system of $a$,
\begin{eqnarray}
\begin{aligned}
 F_{L=0}(q)&=1,\\
 F_{L=1}(q)&=\sqrt{\frac{z_0^2+1}{z^2+1}},\\
 F_{L=2}(q)&=\sqrt{\frac{z_0^4+3z_0^2+9}{z^4+3z^2+9}}\,,
\end{aligned}
\end{eqnarray}
where $z=qR$ and $z_0=q_0R$. Here $q_0$ represents the values of $q$, when the invariant mass is
equal to the nominal mass of the resonance. The effective radius of the barrier $R$ is fixed
to 3.0~GeV$^{-1}$ for the intermediate resonances and 5.0~GeV$^{-1}$ for the
$D_s^+$ meson.

\subsubsection{Propagator}
The intermediate resonances $a_0(1450)$, $a_2(1320)$, $\pi_1(1400)$ and $\pi_1(1600)$ are
parameterized as relativistic Breit-Wigner functions,
\begin{eqnarray}\begin{aligned}
  \begin{array}{lr}
    P = \frac{1}{m_{0}^{2} - s_{a} - im_{0}\Gamma(m)}\,, &\\
    \Gamma(m) = \Gamma_{0}\left(\frac{q}{q_{0}}\right)^{2L+1}\Big(\frac{m_{0}}{m}\Big)\left(\frac{F_{L}(q)}{F_{L}(q_{0})}\right)^{2}\,, &
  \end{array}\label{RBW}
\end{aligned}\end{eqnarray}
where $s_{a}$ denotes the invariant-mass squared of the parent particle;
$m_{0}$ and $\Gamma_{0}$ are the nominal mass and width of each intermediate
resonance, respectively.

We parameterize the $\rho^{0}$ resonance by the Gounaris-Sakurai lineshape~\cite{PhysRevLett.21.244}, which is given by
\begin{eqnarray}\begin{aligned}
P_{\rm GS}(m)=\frac{1+d\frac{\Gamma_0}{m_0}}{m_0^2-m^2+f(m)-im_0\Gamma(m)}\,,
\end{aligned}\end{eqnarray}
where 
\begin{eqnarray}\begin{aligned}
d=\frac{3m^2_\pi}{\pi q_0^2}\ln\left(\frac{m_0+2q_0}{2m_\pi}\right)+\frac{m_0}{2\pi q_0}-\frac{m^2_\pi m_0}{\pi q^3_0}\,.
\end{aligned}\end{eqnarray}
The function $f(m)$ is given by
\begin{eqnarray}
\begin{aligned}
f(m)=\Gamma_0\frac{m_0^2}{q_0^3}\left[q^2(h(m)-h(m_0))+(m_0^2-m^2)q_0^2\left.\frac{dh}{d(m^2)}\right|_{m_0^2}\right]\,,
\end{aligned}
\end{eqnarray}
where
\begin{eqnarray}\begin{aligned}
h(m)=\frac{2q}{\pi m}\ln\left(\frac{m+2q}{2m_{\pi}}\right)\,
\end{aligned}\end{eqnarray}
and
\begin{eqnarray}\begin{aligned}
&\left.\frac{dh}{d(m^2)}\right|_{m_0^2}=
&h(m_0)\left[(8q_0^2)^{-1}-(2m_0^2)^{-1}\right]+(2\pi m_0^2)^{-1}\,.
\end{aligned}\end{eqnarray}

\subsubsection{Spin factors}
The spin-projection operators are defined as~\cite{covariant-tensors}
\begin{eqnarray}
\begin{aligned}
  P^{(1)}_{\mu\mu^{\prime}}(a) &= -g_{\mu\mu^{\prime}}+\frac{p_{a,\mu}p_{a,\mu^{\prime}}}{p_{a}^{2}}\,,\\
  P^{(2)}_{\mu\nu\mu^{\prime}\nu^{\prime}}(a) &= \frac{1}{2}(P^{(1)}_{\mu\mu^{\prime}}(a)P^{(1)}_{\nu\nu^{\prime}}(a)+P^{(1)}_{\mu\nu^{\prime}}(a)P^{(1)}_{\nu\mu^{\prime}}(a))\\
  &-\frac{1}{3}P^{(1)}_{\mu\nu}(a)P^{(1)}_{\mu^{\prime}\nu^{\prime}}(a)\,.
 \label{spin-projection-operators}
\end{aligned}
\end{eqnarray}
The quantities $p_a$, $p_b$, and $p_c$ are the momenta of particles $a$,
$b$, and $c$, respectively, and $r_a = p_b-p_c$.
The covariant tensors are given by
\begin{eqnarray}
\begin{aligned}
    \tilde{t}^{(1)}_{\mu}(a) &= -P^{(1)}_{\mu\mu^{\prime}}(a)r^{\mu^{\prime}}_{a}\,,\\
    \tilde{t}^{(2)}_{\mu\nu}(a) &= P^{(2)}_{\mu\nu\mu^{\prime}\nu^{\prime}}(a)r^{\mu{\prime}}_{a}r^{\nu^{\prime}}_{a}\,.\\
\label{covariant-tensors}
\end{aligned}
\end{eqnarray}
The spin factors for $S$, $P$, and $D$ wave decays are
\begin{eqnarray}
\begin{aligned}
    S &= 1\,, &(S\ \text{wave}), &\\
    S &= \tilde{T}^{(1)\mu}(D_{s}^{\pm})\tilde{t}^{(1)}_{\mu}(a)\,,         &(P\ \text{wave}),\\
    S &= \tilde{T}^{(2)\mu\nu}(D_{s}^{\pm})\tilde{t}^{(2)}_{\mu\nu}(a)\,,         &(D\ \text{wave}),
\label{spin-factor}
\end{aligned}
\end{eqnarray}
where the $\tilde{T}^{(l)}$ factors have the same definitions as $\tilde{t}^{(l)}$. The
tensor describing the $D_{s}^{+}$ decay is denoted by $\tilde{T}$ and
that of the $a$ decay is denoted by $\tilde{t}$.

\subsection{Fit results}
\begin{figure}[htbp]
	\centering 
	\subfigure{\includegraphics[width=0.35\textwidth]{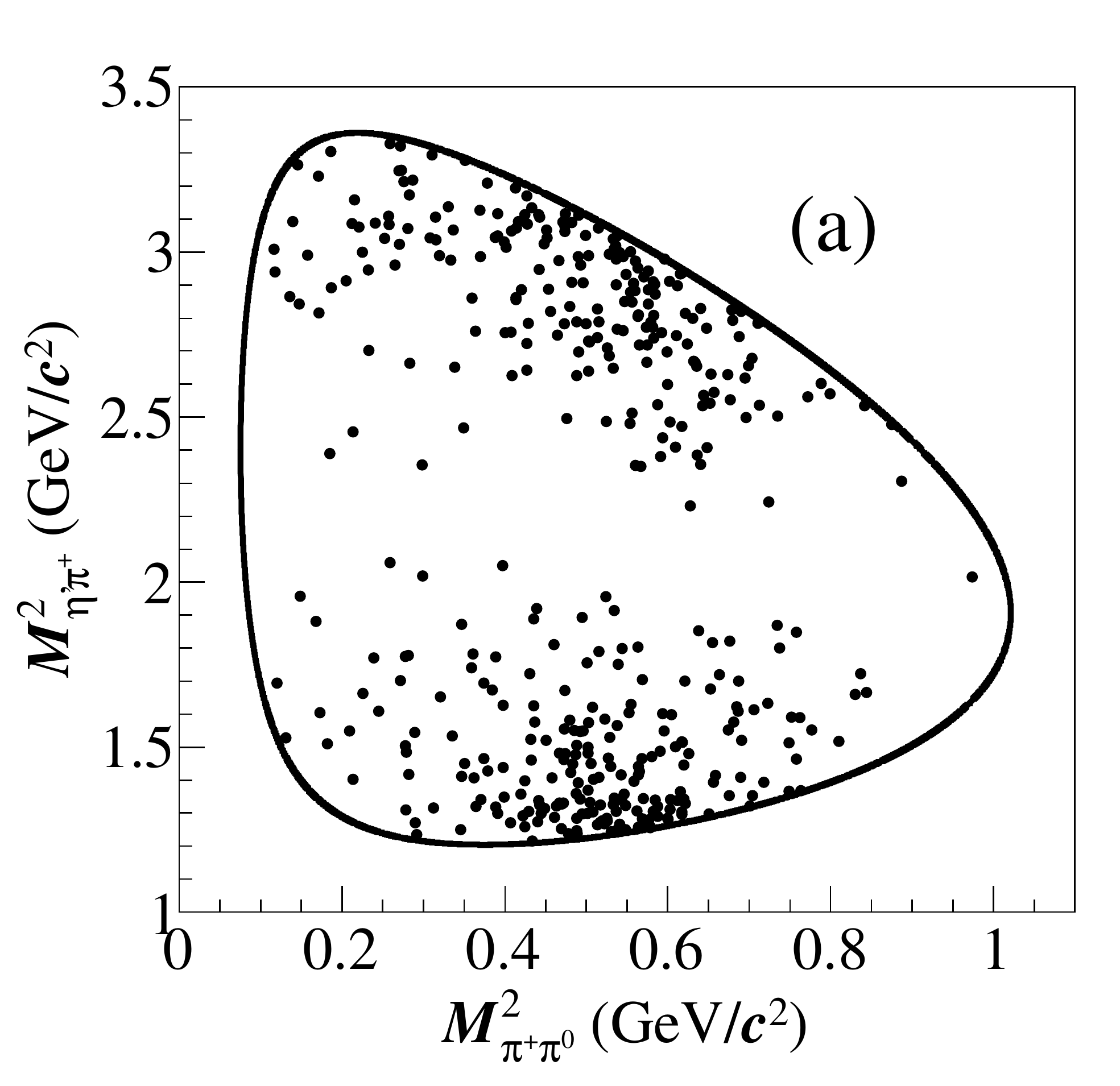}}
	\subfigure{\includegraphics[width=0.35\textwidth]{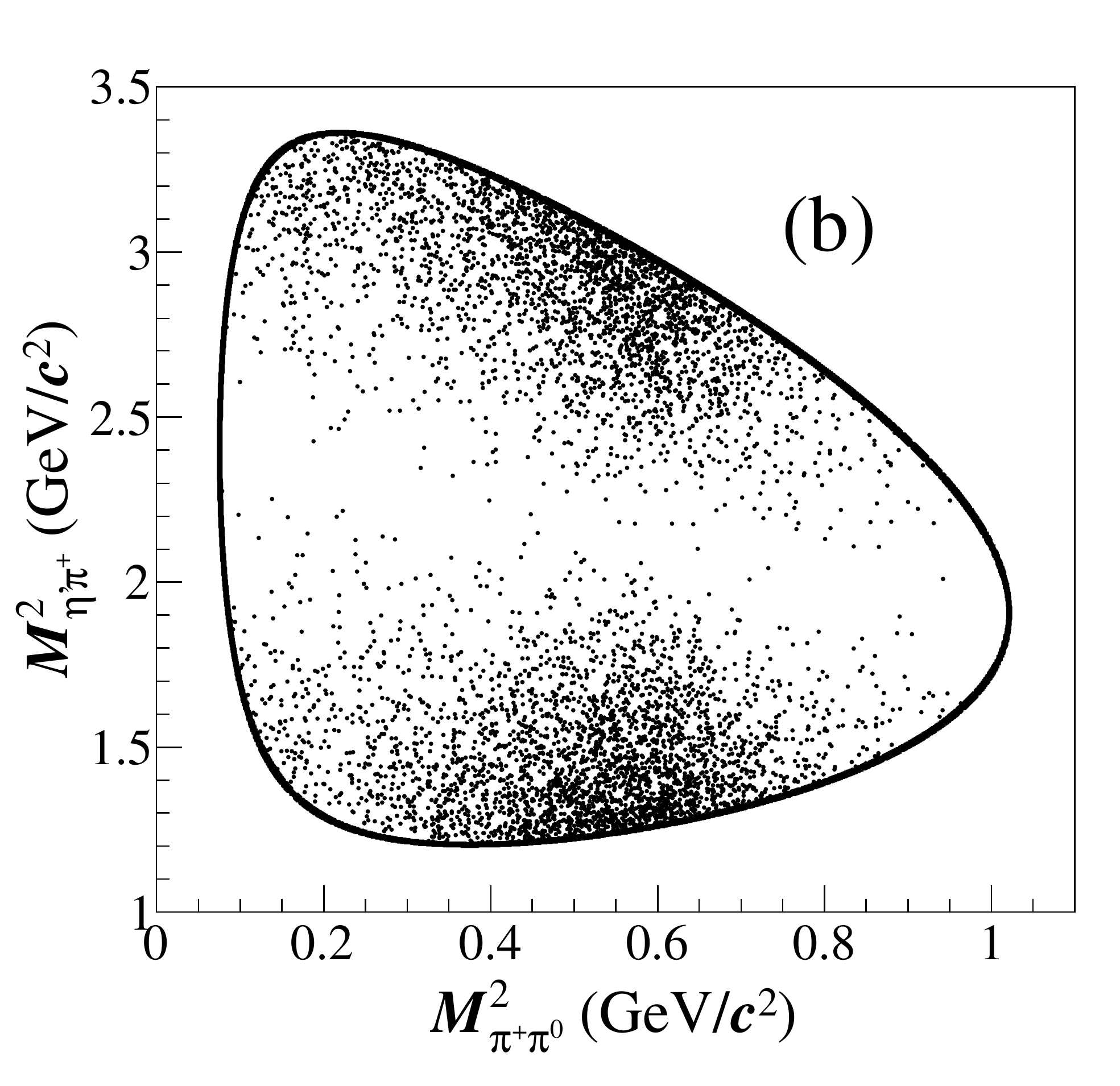}}
	\caption{The Dalitz plots of $m^{2}_{\eta^{\prime}\pi^+}$ versus $m^{2}_{\pi^+\pi^0}$ for (a) the data sample and (b) the signal MC samples generated based on the results of the amplitude analysis at $\sqrt{s} = 4.178-4.226$ GeV. The physical border is indicated by the black line.}
	\label{fig:dalitz}
\end{figure}

\begin{figure}[htbp]
	\centering
	\subfigure{
		\includegraphics[width=0.3\textwidth]{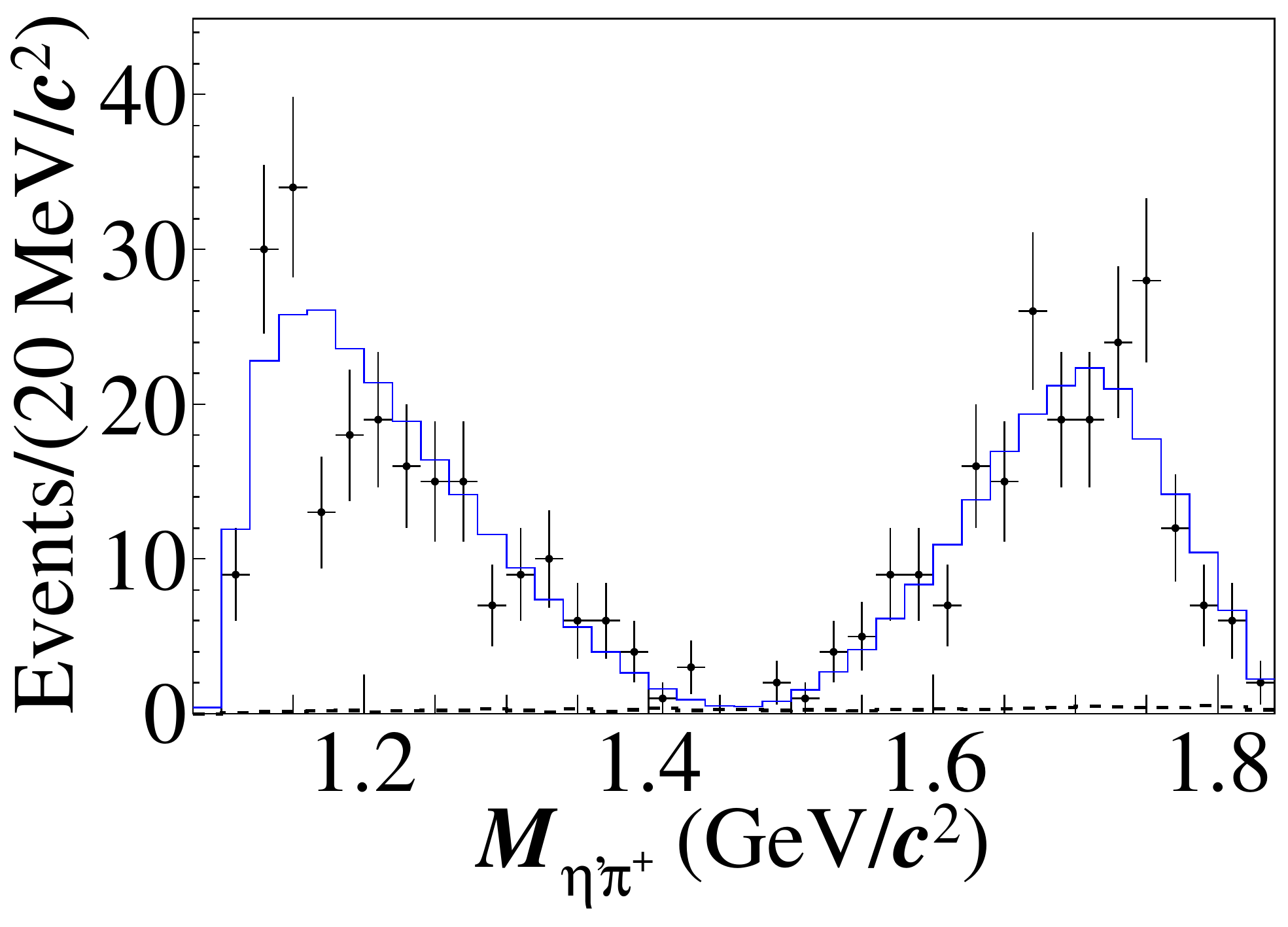}}
	\subfigure{
		\includegraphics[width=0.3\textwidth]{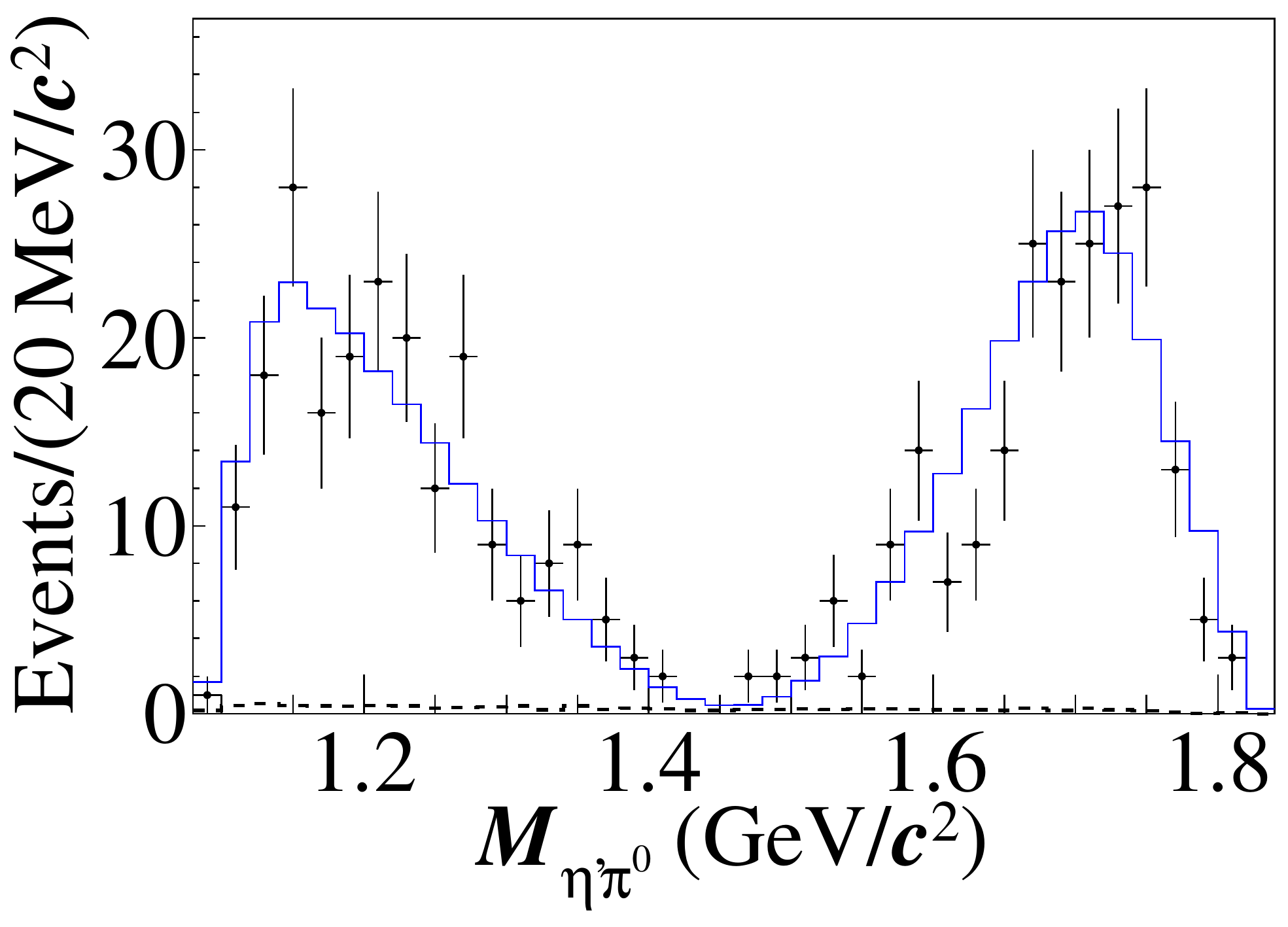}}
	\subfigure{       
		\includegraphics[width=0.3\textwidth]{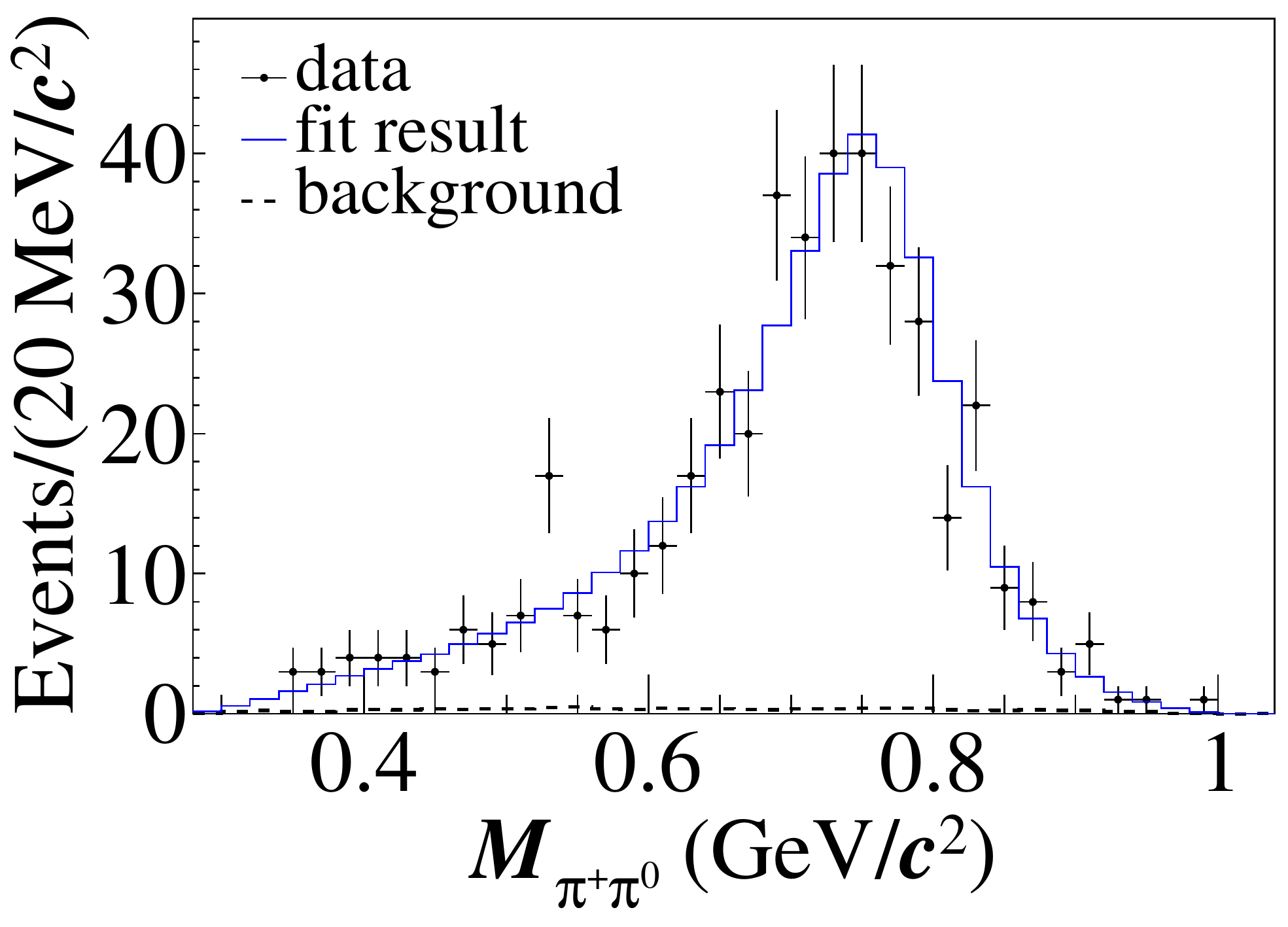}}
	\caption{Projections on $M_{\eta^{\prime}\pi^+}$ (left), $M_{\eta^{\prime}\pi^0}$ (middle) and $M_{\pi^+\pi^0}$ (right) of the nominal fit. The data are represented by points with error bars, the fit results by the blue solid line, and the background events of signal-subtracted inclusive MC sample by the black dashed line.}
	\label{fig:projectoinAmp}
\end{figure}

The Dalitz plots of $M^{2}_{\pi^+\pi^0}$ versus~$M^{2}_{\eta^{\prime}\pi^+}$ for the data samples and the signal MC samples generated based on the results of the amplitude analysis are shown in Fig.~\ref{fig:dalitz}~(a) and Fig.~\ref{fig:dalitz}~(b), respectively. One can see a clear $\rho^{+}$ resonance. Therefore we choose the $D^+_s\to\rho^{+}\eta^{\prime}$ amplitude as a reference, and fix the magnitude and the phase of its amplitude to 1.0 and 0.0, respectively, while those of other amplitudes are floated. The masses and widths of all resonances are fixed to the corresponding PDG averages~\cite{Zyla:2020zbs}, and $w_{\rm sig}$ are fixed to the purities discussed in Sec.~\ref{AASelection}. Then we test other possible intermediate resonances, such as $\rho(1450)$, $a_0(1450)$, $\pi_1(1600)$, $a_2(1320)$, etc., by adding them one by one. We also examined the possible combinations of these intermediate resonances to check their significances, correlations and interferences. We use the difference of log-likelihoods of fits with and without these amplitudes to calculate the significance and find that in all cases these significances are less than three standard deviations. The significance of each intermediate resonance tested is listed in Table~\ref{tab:testAmp}. Hence the final model consists only of the mode $D^+_s\to\rho^{+}\eta^{\prime}$. The mass projections of the fit results are shown in Fig.~\ref{fig:projectoinAmp}.

\begin{figure}[htbp]
	\centering
	\subfigure{
		\includegraphics[width=0.4\textwidth]{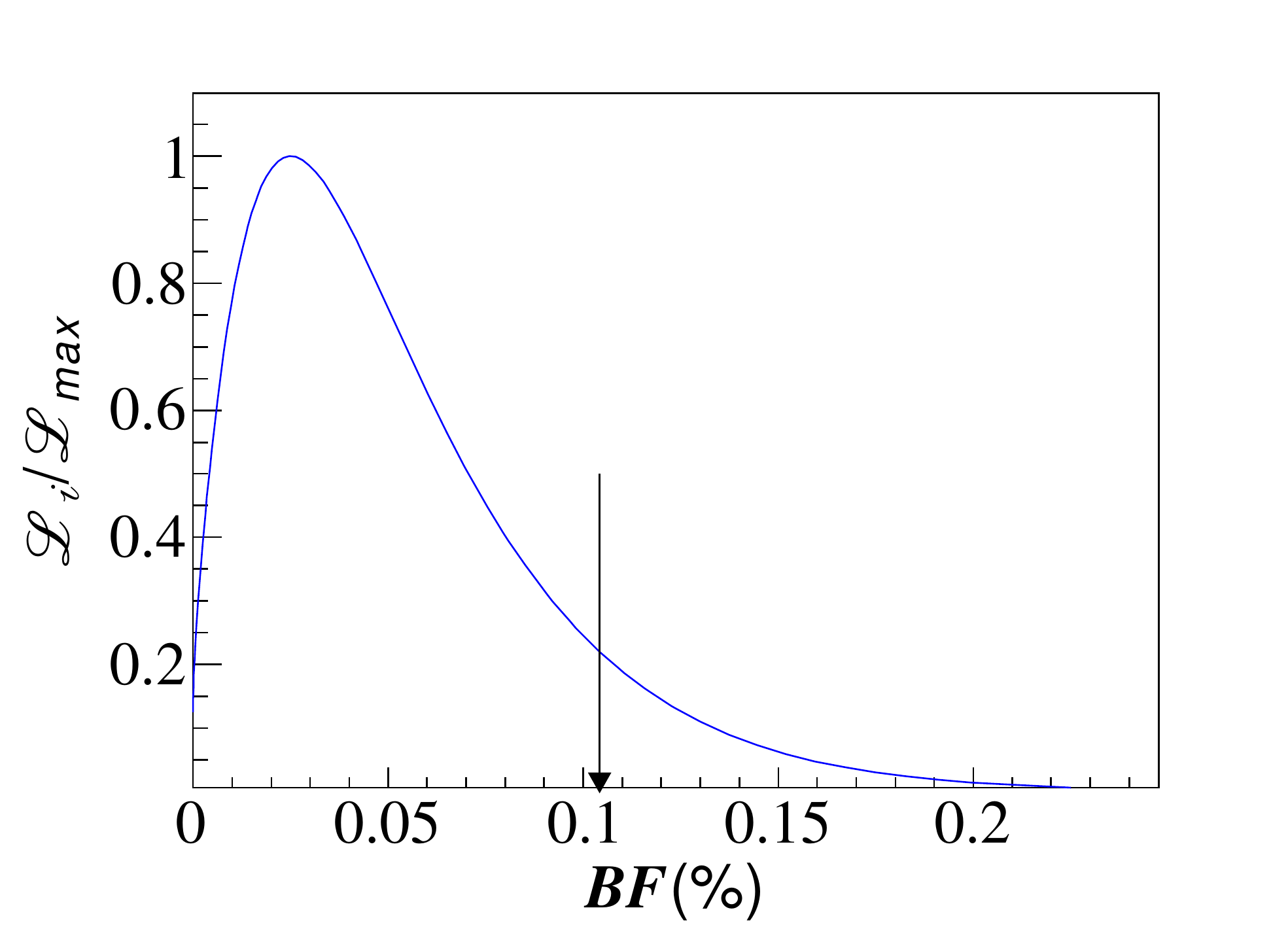}}
	\subfigure{
		\includegraphics[width=0.4\textwidth]{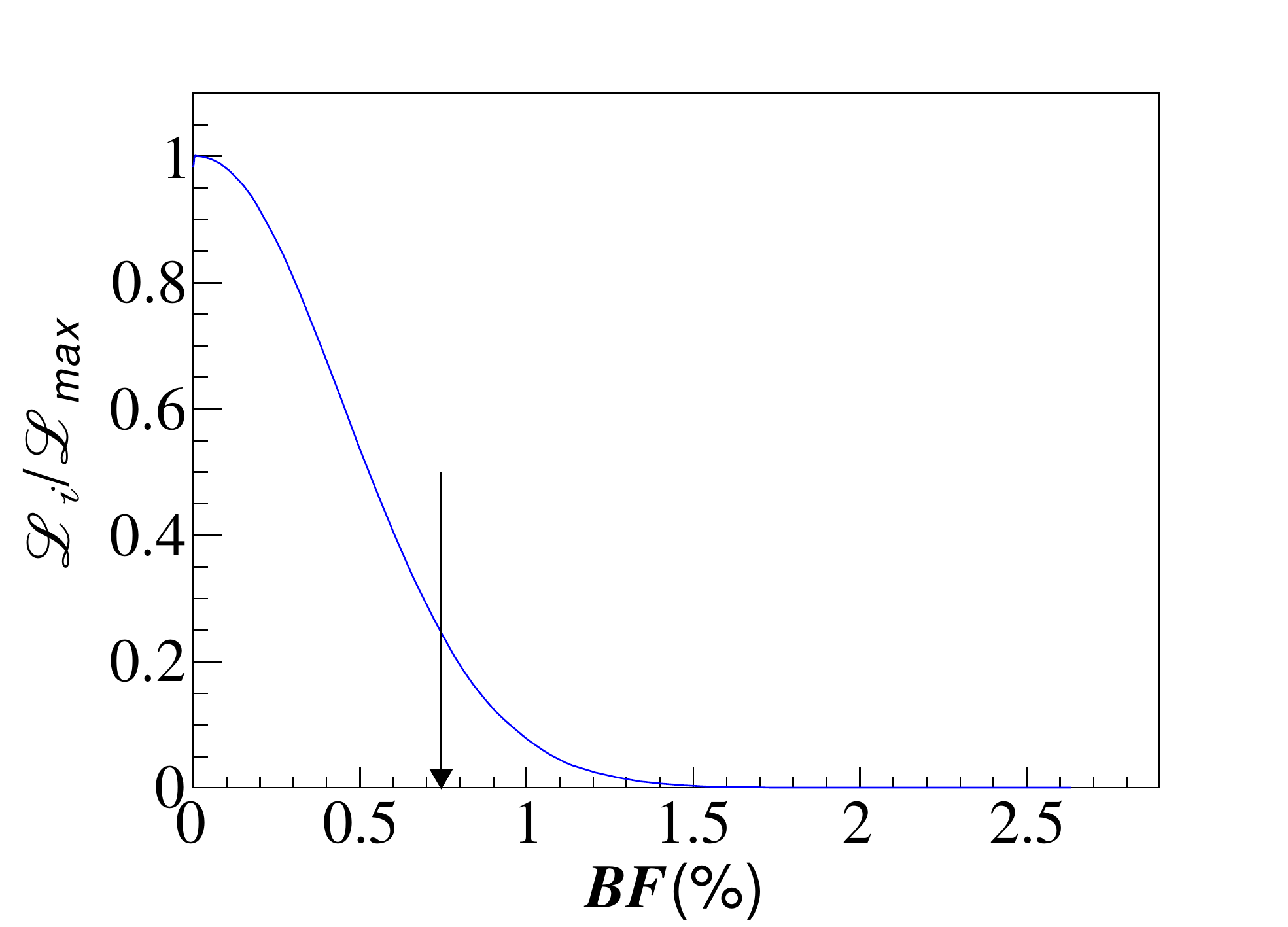}}
	\caption{Likelihood versus the BF of $D^+_s\to(\pi^+\pi^0)_{S}\eta^{\prime}$ (left) and $D^+_s\to(\pi^+\pi^0)_{P}\eta^{\prime}$ (right). The result obtained with incorporating the systematic uncertainty is shown in blue curve. The $\mathscr{L}_{max}$ denotes the maximum likelihood obtained from the fit. The black arrows show the results corresponding to the 90\% confidence level.
	}
	\label{fig:FF-up}
\end{figure}

In addition, we also try including the $S$-wave and $P$-wave nonresonant components, which are denoted as $D^+_s\to (\pi^+\pi^0)_{S}\eta^{\prime}$ and $D^+_s\to (\pi^+\pi^0)_{P}\eta^{\prime}$, respectively. The significances of the nonresonant processes are both less than three standard deviations. For the BFs of nonresonant decays, we scan the magnitudes of the nonresonant decays to obtain the likelihood variation versus the expected BF as shown in Fig.~\ref{fig:FF-up}. To take the uncertainty of total BF shown in Table~\ref{tab:bfsys} and systematic uncertainty of amplitude analysis listed in Table~\ref{tab:pwaSys} into account, the likelihood is convolved with a Gaussian function with a width equal to the total systematic uncertainty. The total systematic uncertainties of $S$-wave and $P$-wave components are 2.9\% and 4.4\%, respectively. Finally, we obtain the upper limits $\mathcal{B}(D^+_s\to (\pi^+\pi^0)_{S}\eta^{\prime}) < 0.10\%$ and $\mathcal{B}(D^+_s\to(\pi^+\pi^0)_{P}\eta^{\prime}) < 0.74\%$ at the 90\% confidence level.
\begin{table}[htbp]
	\caption{Tested amplitudes, but not included in the nominal fit. }
	\centering
	\begin{tabular}{lcc}
		\hline
		{Amplitude}   &{significance ($\sigma$)}\\
		\hline
		$D^+_s\to\rho(1450)^+\eta^{\prime}, \rho(1450)^{+}\to\pi^+\pi^0$     &0.9 \\
		$D^+_s\to a_0(1450)^+\pi^0, a_0(1450)^+\to\pi^+\eta^{\prime}$        &2.4 \\     
		$D^+_s\to a_0(1450)^0\pi^+, a_0(1450)^0\to\pi^0\eta^{\prime}$        &2.8 \\
		$D^+_s\to \pi_1(1400)^+\pi^0, \pi_1(1400)^+\to\pi^+\eta^{\prime}$    &0.5 \\
		$D^+_s\to \pi_1(1400)^0\pi^+, \pi_1(1400)^0\to\pi^0\eta^{\prime}$    &1.4 \\
		$D^+_s\to \pi_1(1600)^+\pi^0, \pi_1(1600)^+\to\pi^+\eta^{\prime}$    &2.7 \\
		$D^+_s\to \pi_1(1600)^0\pi^+, \pi_1(1600)^0\to\pi^0\eta^{\prime}$    &2.2 \\
		$D^+_s\to a_2(1320)^+\pi^0, a_2(1320)^+\to\pi^+\eta^{\prime}$       &0.3 \\
		$D^+_s\to a_2(1320)^0\pi^+, a_2(1320)^0\to\pi^0\eta^{\prime}$        &2.6 \\
		\hline
	\end{tabular}
	\label{tab:testAmp}
\end{table}
\subsection{Systematic uncertainties for amplitude analysis}
\label{sec:PWA-Sys}
The following four sources of potential bias are considered when assigning systematic uncertainties.
\begin{itemize}
\item[\lowercase\expandafter{\romannumeral1}]
  Resonance parameters. The uncertainties related to the fixed parameters in the amplitudes are estimated by varying the masse and width of the $\rho^+$ resonance by $\pm 1 \sigma$ \cite{Zyla:2020zbs}.

\item[\lowercase\expandafter{\romannumeral2}]
The $\rho^+$ lineshape. The uncertainties related to the lineshape of the $\rho^+$ are estimated by using a Breit-Wigner function instead of the Gounaris-Sakurai description.

\item[\lowercase\expandafter{\romannumeral3}]
  $R$ values. The radii of the nonresonant states and $D_s^{\pm}$ mesons
  are varied within the range $[2.0, 4.0]$~GeV$^{-1}$ for intermediate
  resonances and $[4.0, 6.0]$~GeV$^{-1}$ for $D_s^{\pm}$ mesons.

\item[\lowercase\expandafter{\romannumeral4}]
  Background estimation. The uncertainties associated with the background estimation are studied by varying the
  signal fraction, i.e.~$w_{\rm sig}$ in Eq.~(\ref{likelihood3}), by its
  statistical uncertainty. The largest differences from the nominal results
  are assigned as the uncertainties. The other source of potential bias arise from the knowledge of the
  background distributions. An alternative MC-simulated shape is used where the relative fractions of backgrounds from $q\bar{q}$ and non-$D_{s}^{*\pm}D_{s}^{\mp}$ open charm are varied by the statistical uncertainties of their cross sections.  
  
\item[\lowercase\expandafter{\romannumeral5}]
   Detector effects. These effects are related to the efficiency difference between MC simulation and data caused by PID and tracking, reflected in the $\gamma_{\epsilon}(p)$ in Eq.(~\ref{eq::gamma}). The uncertainties associated with $\gamma_{\epsilon}(p)$ are obtained by performing alternative amplitude analyses varying PID and tracking efficiencies according to their uncertainties. The systematic uncertainty from this source is found to be negligible.
\end{itemize}

The assigned systematic uncertainties on the fit fractions (FF) for the $S$-wave and $P$-wave nonresonant components are summarized in Table~\ref{tab:pwaSys}. The FF for the $L$-wave nonresonant amplitude is defined as
\begin{eqnarray}\begin{aligned}
{\rm FF}_{L} = \frac{\sum^{N_{\rm gen}} \left|\rho_{L}\mathcal{N}_{L}\right|^{2}}{\sum^{N_{\rm gen}} \left|\mathcal{M}\right|^{2}}\,, \label{Fit-Fraction-Definition}
\end{aligned}\end{eqnarray}
where $N_{\rm gen}$ is the number of phase-space MC events at generator level. It involves the phase-space MC truth information without detector acceptance or resolution effects.
\begin{table}[htbp]
	\caption{Systematic uncertainties on FFs for  $S$-wave and $P$-wave nonresonant components.}
	\centering
	\begin{tabular}{c|c|c}
		\hline
		\hline
		\multirow{2}{*}{Source} & \multicolumn{2}{c}{Systematic Uncertainty(\%)} \\
		\cline{2-3}
		& $D^+_s\to (\pi^+\pi^0)_{S}\eta^{\prime}$ &$D^+_s\to (\pi^+\pi^0)_{P}\eta^{\prime}$\\
		\hline
		Resonance parameters     &<0.1  &<0.1 \\
		$\rho^+$ lineshape &<0.1  &<0.1 \\
		$R$ values               &0.1  &3.3\\
		Background               &0.2  &0.1\\
		\hline
		Total                    &0.2  &3.3\\
		\hline
		\hline
	\end{tabular}
	\label{tab:pwaSys}
\end{table}

\section{Branching fraction measurement}
\label{BFSelection}
\begin{figure*}[htp]
\begin{center}
\includegraphics[width=0.90\textwidth]{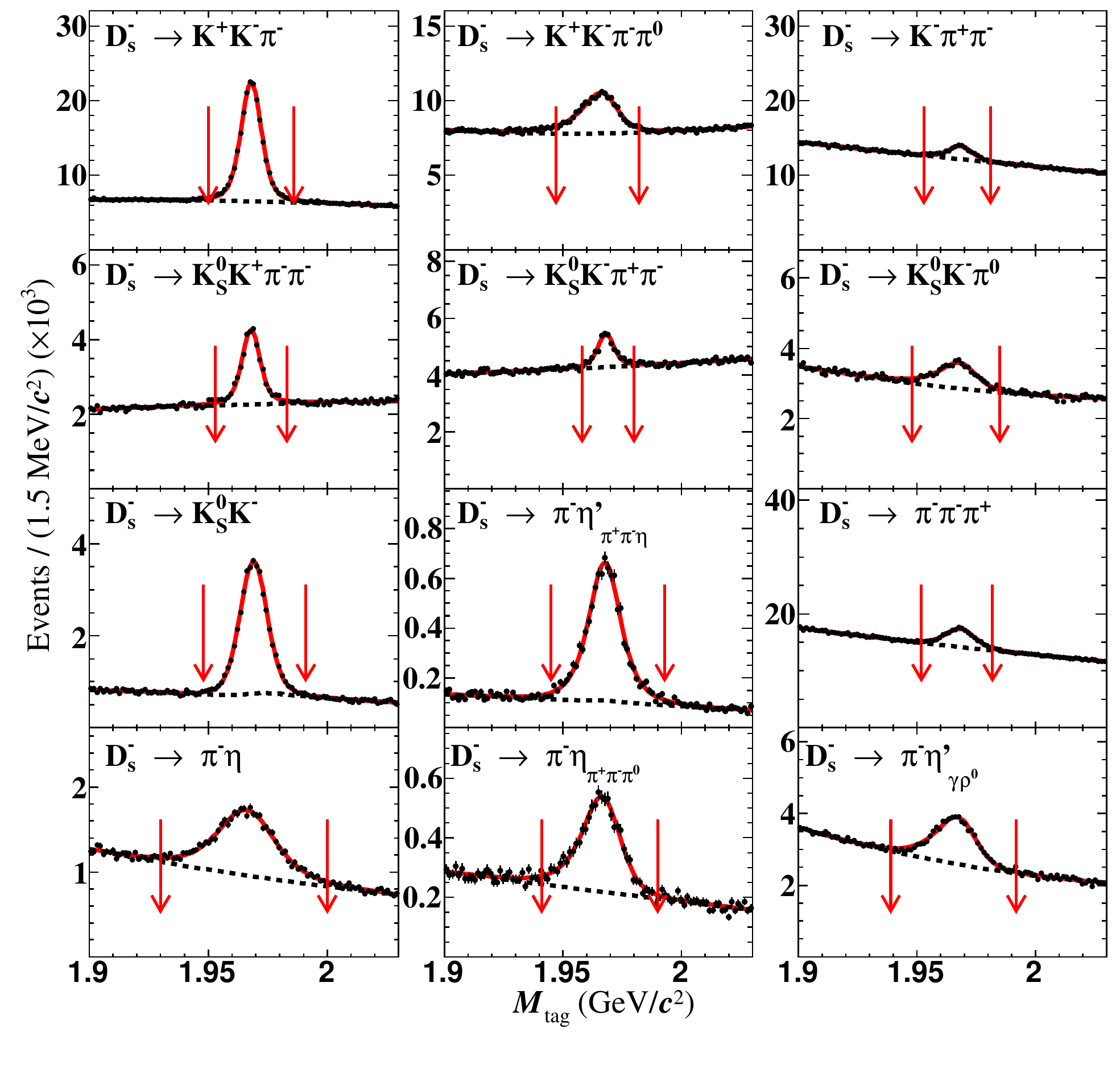}
\caption{Fits to the $M_{\rm tag}$ distributions of the ST candidates
         from the data sample at $\sqrt{s}=4.178$~GeV. The points with
         error bars are data, the red solid lines are the total fits, and the black
         dashed lines are background. The pairs of red arrows denote the
         signal regions.
         }
\label{fit:Mass-data-Ds_4180}
\end{center}
\end{figure*}

The ST $D_{s}^{-}$ mesons are reconstructed through all twelve hadronic decays as shown in Table~\ref{tab:tag-cut} and the selection criteria are the same as those described in Sec.~\ref{ST-selection} for the branching fraction measurement. In addition, all pions are required to have momenta greater than 100~MeV/$c$ to remove soft pions from $D^{*+}$ decays. The best tag candidate with $M_{\rm rec}$ closest to the $D_{s}^{*\pm}$ known mass~\cite{Zyla:2020zbs} is chosen if there are multiple ST candidates. The data sets are organized into three sample groups, 4.178~GeV, 4.189-4.219~GeV, and 4.226~GeV, that were acquired during the same year under consistent running condition. The yields for various tag modes are obtained by fitting the corresponding $M_{\rm tag}$ distributions. As an example, the fits to the $M_{\rm tag}$ distributions of the accepted ST candidates from the data sample at $\sqrt s=4.178$~GeV are shown in Fig.~\ref{fit:Mass-data-Ds_4180}.
In the fits, the signal is modeled by a MC-simulated shape convolved with a Gaussian function to account for differences in resolution between data and MC simulation. The background is described by a second-order Chebyshev polynomial. Inclusive MC studies show that there is no peaking background in any tag mode, except for
$D^{-} \to K_{S}^{0} \pi^-$ and $D_{s}^{-} \to \eta\pi^+\pi^-\pi^-$ faking the
$D_{s}^{-} \to K_{S}^{0} K^-$ and $D_{s}^{-} \to \pi^-\eta^{\prime}$ tags,
respectively. Therefore, the MC-simulated shapes of these two
peaking background sources are added to the background polynomial functions.

Once a tag mode is reconstructed, we select the signal decay
$D_{s}^{+} \to \pi^{+}\pi^{0}\eta^{\prime}$. In the case of multiple candidates, the
DT candidate with the average mass, $(M_{\rm sig}+M_{\rm tag})/2$,
closest to the $D_{s}^{\pm}$ nominal mass listed in the PDG~\cite{Zyla:2020zbs} is retained.

To measure the BF, we employ the following equations:
\begin{eqnarray}\begin{aligned}
  N_{\text{tag}}^{\text{ST}} = 2N_{D_{s}^{+}D_{s}^{-}}\mathcal{B}_{\text{tag}}\epsilon_{\text{tag}}^{\text{ST}}\,, \label{eq-ST}
\end{aligned}\end{eqnarray}
\begin{equation}
  N_{\text{tag,sig}}^{\text{DT}}=2N_{D_{s}^{+}D_{s}^{-}}\mathcal{B}_{\text{tag}}\mathcal{B}_{\text{sig}}\epsilon_{\text{tag,sig}}^{\text{DT}}\,,
  \label{eq-DT}
\end{equation}
where $N_{\text{tag}}^{\text{ST}}$ is the ST yield for the tag mode; $N_{\text{tag,sig}}^{\text{DT}}$ is the DT yield; $N_{D_{s}^{+}D_{s}^{-}}$ is the total number of $D_{s}^{*\pm}D_{s}^{\mp}$
pairs produced from the $e^{+}e^{-}$ collisions; $\mathcal{B}_{\text{tag}}$ and $\mathcal{B}_{\text{sig}}$ are the BFs of the
tag and signal modes, respectively; $\epsilon_{\text{tag}}^{\text{ST}}$ is the
ST efficiency to reconstruct the tag mode; and $\epsilon_{\text{tag,sig}}^{\text{DT}}$
is the DT efficiency to reconstruct both the tag and the signal decay modes. In
the case of more than one tag mode and sample group,
\begin{eqnarray}
\begin{aligned}
  \begin{array}{lr}
    N_{\text{total}}^{\text{DT}}=\Sigma_{\alpha, i}N_{\alpha,\text{sig},i}^{\text{DT}}   = \mathcal{B}_{\text{sig}}
 \Sigma_{\alpha, i}2N^{i}_{D_{s}^{+}D_{s}^{-}}\mathcal{B}_{\alpha}\epsilon_{\alpha,\text{sig}, i}^{\text{DT}}\,,
  \end{array}
  \label{eq-DTtotal}
\end{aligned}
\end{eqnarray}
where $\alpha$ represents the  tag mode in the $i^{\rm th}$ sample group.
By isolating $\mathcal{B}_{\text{sig}}$ and replacing $N_{D_{s}^{+}D_{s}^{-}}$ shown in Eq.~(\ref{eq-ST}) , we find
\begin{eqnarray}\begin{aligned}
  \label{equ:BF}
  \mathcal{B}_{\text{sig}} =
  \frac{N^{\text{DT}}_{\text{fitted}}-N^{\text{DT}}_{\text{peaking}}}{ \mathcal{B}_{\eta \to \gamma\gamma}\mathcal{B}_{\pi^0 \to \gamma\gamma}\mathcal{B}_{\eta^{\prime} \to \pi^+\pi^-\eta}\begin{matrix}\sum_{\alpha, i} N_{\alpha, i}^{\text{ST}}\epsilon^{\text{DT}}_{\alpha,\text{sig},i}/\epsilon_{\alpha,i}^{\text{ST}}\end{matrix}}\, , 
\end{aligned}\end{eqnarray}
where $N_{\alpha,i}^{\text{ST}}$ is obtained from the data sample, while $N^{\text{DT}}_{\text{peaking}}$, $\epsilon_{\alpha,\text{sig},i}^{\text{DT}}$ and $\epsilon_{\alpha,i}^{\text{ST}}$ are
obtained from the inclusive MC sample. The $D_{s}^{+} \to \pi^{+}\pi^{0}\eta^{\prime}$ simulated sample is generated
according to the results of the amplitude analysis. The BFs $\mathcal{B}_{\eta \to \gamma\gamma}$, $\mathcal{B}_{\pi^0 \to \gamma\gamma}$ and $\mathcal{B}_{\eta^{\prime} \to \pi^+\pi^-\eta}$ have been introduced to consider these sub-channels.

\begin{figure}[!htbp]
  \centering
  \includegraphics[width=0.45\textwidth]{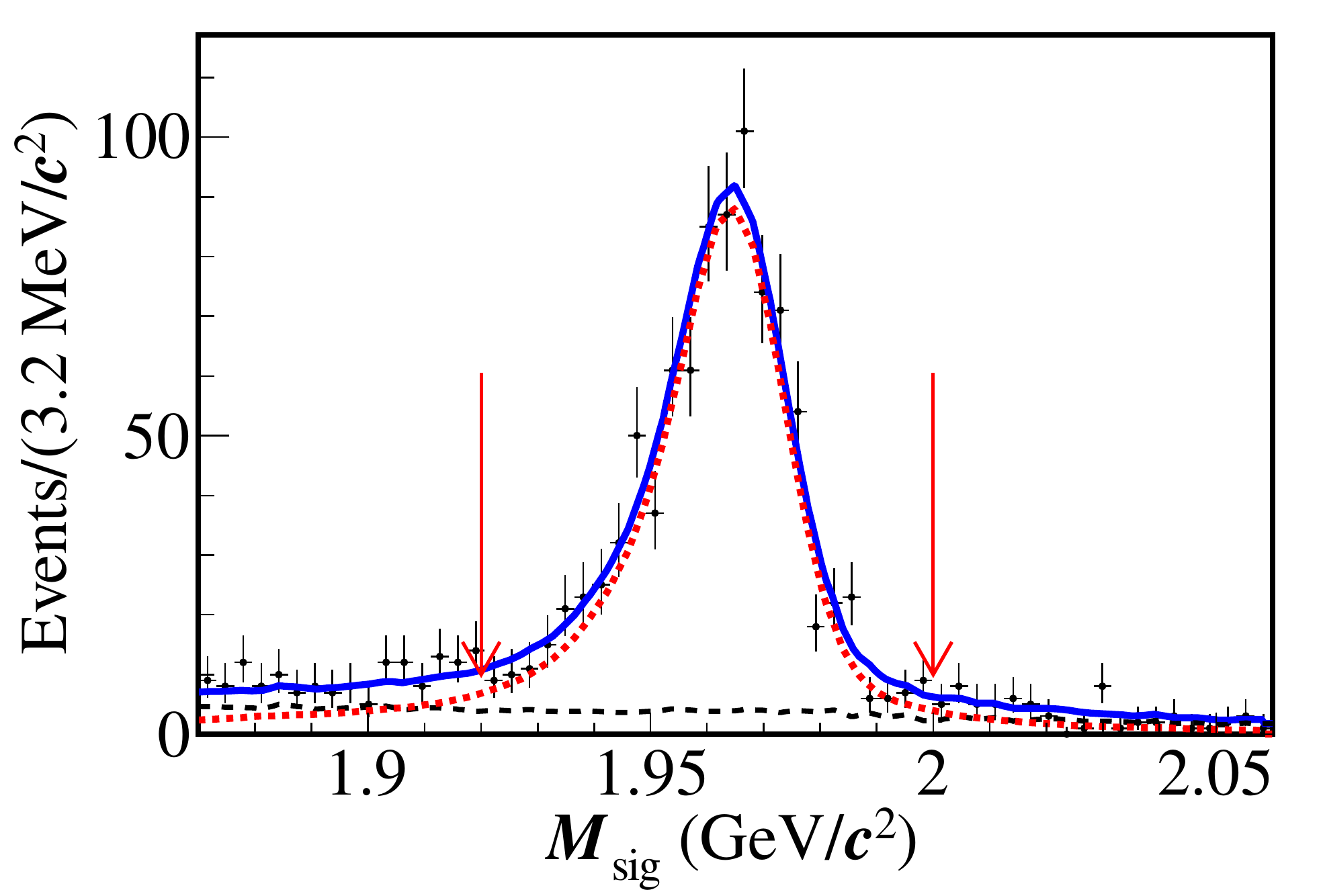}
 \caption{Fit to the $M_{\rm sig}$ distribution of the DT candidates from the
   data samples at $\sqrt{s}= 4.178$-$4.226$~GeV. This plot is obtained using twelve tag modes in Table~\ref{tab:tag-cut}. The data are represented by points with error bars, the total fit by the blue solid line, and the fitted
   signal and the fitted background by the red dotted and the black dashed
   lines, respectively. The pair of red arrows indicate the signal region.
 }
  \label{DT-fit}
\end{figure}

The $N^{\text{DT}}_{\text{fitted}}$ is obtained from the fit to the $M_{\rm sig}$ distribution of the selected $D^+_s\to \pi^+\pi^0\eta^{\prime}$ candidates. The fit result is shown in Fig.~\ref{DT-fit}, where the signal shape is described by a MC-simulated shape convolved with a Gaussian function to account for differences in resolution between data and MC. The background shape is described by a MC-simulated shape which excludes peaking background from $D^+_s \to \pi^+ \eta \omega_{\pi^{+}\pi^{-}\pi^{0}}$. The number of  peaking background events, $N^{\text{DT}}_{\text{peaking}}$  is estimated from the inclusive MC sample. Thus, $N^{\text{DT}}_{\text{fitted}}$ and $N^{\text{DT}}_{\text{peaking}}$  are determined to be $837\pm35$ and $5\pm1$, respectively. Tables~\ref{ST-eff1} -~\ref{ST-eff3} summarize the ST efficiencies, DT efficiencies, and ST yields in data samples at the C.M. energies $\sqrt{s} = 4.178-4.226$ GeV. Taking into account the differences in $\pi^{\pm}$ tracking/PID efficiencies, $\pi^{0}$ and $\eta$ reconstruction efficiencies between data and MC simulation, we determine the BF $\mathcal{B}(D^+_s\to\pi^+\pi^0 \eta^{\prime})=(6.15\pm0.25(\rm stat.)\pm0.18(\rm syst.))\%$ according to Eq.~(\ref{equ:BF}).

\begin{table}[htbp]
	\caption{The efficiencies and ST yields at $\sqrt{s} = 4.178$~GeV. The uncertainties are statistical.} 
	\centering
	\begin{tabular}{lccc}
		\hline
        Tag mode                                              &$N_{\rm ST}$                   &$\epsilon_{ST}(\%)$        & $\epsilon_{DT}(\%)$ \\
		\hline
		$D_{s}^{-}\to K_{S}^{0}K^{-}$                         & $\phantom{0}31941\pm312$      &$47.36\pm0.07$             &$6.30\pm0.18$\\  
		$D_{s}^{-}\to K^{+}K^{-}\pi^{-}$                      & $137240\pm614$                &$39.47\pm0.03$             &$5.08\pm0.07$\\
		$D_{s}^{-}\to K_{S}^{0}K^{-}\pi^{0}$                  & $\phantom{0}11385\pm529$      &$16.12\pm0.11$             &$2.19\pm0.10$\\		
		$D_{s}^{-}\to K_{S}^{0}K^{-}\pi^{-}\pi^{+}$           & $\phantom{00}8093\pm326$      &$20.40\pm0.12$             &$2.06\pm0.13$\\
		$D_{s}^{-}\to K_{S}^{0}K^{+}\pi^{-}\pi^{-}$           & $\phantom{0}15719\pm289$      &$21.83\pm0.06$             &$2.50\pm0.11$\\
		$D^-_s\to \pi^-\eta_{\gamma\gamma}$                   & $\phantom{0}17940\pm402$      &$43.58\pm0.15$             &$5.91\pm0.21$\\
		$D^-_s\to \pi^-\eta^{\prime}_{\pi^+\pi^-\eta_{\gamma\gamma}}$ & $\phantom{00}7759\pm141$      &$19.12\pm0.06$             &$2.22\pm0.13$\\
		$D^-_s\to K^-\pi^+\pi^-$                              & $\phantom{0}17423\pm666$      &$47.46\pm0.22$             &$6.15\pm0.21$\\
		$D^-_s\to K^-K^+\pi^-\pi^0$                           & $\phantom{0}39306\pm799$      &$10.50\pm0.03$             &$1.30\pm0.03$\\
		$D^-_s\to \pi^-\pi^-\pi^+$                            & $\phantom{0}37977\pm859$      &$51.43\pm0.15$             &$6.77\pm0.17$\\		
		$D^-_s\to \pi^-\eta_{\pi^+\pi^-\pi^0}$                & $\phantom{00}5102\pm172$      &$20.85\pm0.10$             &$2.57\pm0.19$\\		
		$D^-_s\to \pi^-\eta^{\prime}_{\gamma\rho^0}$                  & $\phantom{0}20580\pm538$      &$26.28\pm0.10$             &$3.79\pm0.13$\\		
		\hline
	\end{tabular}
	\label{ST-eff1}
\end{table}
\begin{table}[htbp]
	\caption{The efficiencies and ST yields at $\sqrt{s} = 4.189-4.219$~GeV. The uncertainties are statistical.} 
	\centering
	\begin{tabular}{lccc}
		\hline
        Tag mode                                                  &$N_{\rm ST}$                 &$\epsilon_{ST}(\%)$        & $\epsilon_{DT}(\%)$ \\
		\hline
		$D_{s}^{-}\to K_{S}^{0}K^{-}$                             & $18559\pm261$               &$47.26\pm0.09$            &$6.41\pm0.23$\\  
		$D_{s}^{-}\to K^{+}K^{-}\pi^{-}$                          & $81286\pm505$               &$39.32\pm0.04$            &$5.12\pm0.09$\\
		$D_{s}^{-}\to K_{S}^{0}K^{-}\pi^{0}$                      & $\phantom{0}6832\pm457$     &$15.71\pm0.16$            &$2.28\pm0.14$\\
		$D_{s}^{-}\to K_{S}^{0}K^{-}\pi^{-}\pi^{+}$               & $\phantom{0}5269\pm282$     &$20.19\pm0.17$            &$2.13\pm0.17$\\
		$D_{s}^{-}\to K_{S}^{0}K^{+}\pi^{-}\pi^{-}$               & $\phantom{0}8948\pm231$     &$21.63\pm0.09$            &$2.48\pm0.14$\\
		$D^-_s\to \pi^-\eta_{\gamma\gamma}$                       &$10025\pm339$                &$43.00\pm0.22$            &$6.14\pm0.28$\\
		$D^-_s\to \pi^-\eta^{\prime}_{\pi^+\pi^-\eta_{\gamma\gamma}}$     &$\phantom{0}4428\pm111$      &$19.00\pm0.08$            &$2.42\pm0.18$\\
		$D^-_s\to K^-\pi^+\pi^-$                                  &$10175\pm448$                &$47.19\pm0.32$            &$6.22\pm0.28$\\
		$D^-_s\to K^-K^+\pi^-\pi^0$                               &$23311\pm659$                &$10.58\pm0.05$            &$1.24\pm0.04$\\
		$D^-_s\to \pi^-\pi^-\pi^+$                                &$21909\pm776$                &$50.35\pm0.22$            &$7.03\pm0.23$\\		
		$D^-_s\to \pi^-\eta_{\pi^+\pi^-\pi^0}$                    &$\phantom{0}3185\pm146$      &$20.79\pm0.14$            &$2.69\pm0.25$\\		
		$D^-_s\to \pi^-\eta^{\prime}_{\gamma\rho^0}$                      &$11937\pm480$                &$26.09\pm0.14$            &$3.33\pm0.16$\\	
		\hline
	\end{tabular}
	\label{ST-eff2}
\end{table}
\begin{table}[htbp]
	\caption{The efficiencies and ST yields at $\sqrt{s} = 4.226$~GeV. The uncertainties are statistical.} 
	\centering
	\begin{tabular}{lccc}
		\hline
        Tag mode                                                    &$N_{\rm ST}$           &$\epsilon_{ST}(\%)$        & $\epsilon_{DT}(\%)$ \\
		\hline
		$D_{s}^{-}\to K_{S}^{0}K^{-}$                               & $\phantom{0}6582\pm160$             &$46.37\pm0.16$      &$6.41\pm0.37$\\   
		$D_{s}^{-}\to K^{+}K^{-}\pi^{-}$                            & $28439\pm327$                       &$38.38\pm0.07$      &$5.07\pm0.15$\\
		$D_{s}^{-}\to K_{S}^{0}K^{-}\pi^{0}$                        & $\phantom{0}2227\pm220$             &$15.93\pm0.29$      &$1.89\pm0.20$\\	
		$D_{s}^{-}\to K_{S}^{0}K^{-}\pi^{-}\pi^{+}$                 & $\phantom{0}1662\pm217$             &$19.50\pm0.31$      &$2.12\pm0.27$\\
		$D_{s}^{-}\to K_{S}^{0}K^{+}\pi^{-}\pi^{-}$                 & $\phantom{0}3263\pm172$             &$21.29\pm0.15$      &$2.46\pm0.23$\\
		$D^-_s\to \pi^-\eta_{\gamma\gamma}$                         &$\phantom{0}3725\pm252$              &$41.83\pm0.41$      &$6.29\pm0.47$\\
		$D^-_s\to \pi^-\eta^{\prime}_{\pi^+\pi^-\eta_{\gamma\gamma}}$       &$\phantom{0}1648\pm\phantom{0}74$    &$18.56\pm0.13$      &$2.22\pm0.28$\\
		$D^-_s\to K^-\pi^+\pi^-$                                    &$\phantom{0}4984\pm458$              &$45.66\pm0.59$      &$7.14\pm0.50$\\
		$D^-_s\to K^-K^+\pi^-\pi^0$                                 &$\phantom{0}7785\pm453$              &$10.39\pm0.09$      &$1.39\pm0.08$\\
		$D^-_s\to \pi^-\pi^-\pi^+$                                  &$\phantom{0}7511\pm393$              &$49.32\pm0.41$      &$7.06\pm0.37$\\
		$D^-_s\to \pi^-\eta_{\pi^+\pi^-\pi^0}$                      &$\phantom{0}1044\pm\phantom{0}78$    &$20.31\pm0.25$      &$2.63\pm0.40$\\
		$D^-_s\to \pi^-\eta^{\prime}_{\gamma\rho^0}$                        &$\phantom{0}3813\pm335$              &$25.94\pm0.27$      &$3.83\pm0.28$\\
		 
		\hline
	\end{tabular}
	\label{ST-eff3}
\end{table}
The following sources of the systematic uncertainties are taken into account for the BF measurement.

\begin{itemize}
	\item The number of ST $D^{+}_s$ mesons. The systematic uncertainty due to the total yield of the ST $D_s^-$ mesons is assigned to be 0.9\% by taking into account the background fluctuation in the fit, and examining the changes of the fit yields by using alternative signal and background shapes. 
	\item Background shape. The systematic uncertainty due to the MC-simulated background shape is studied by varying the relative fractions of the background from $q\bar{q}$ or non-$D_{s}^{*+}D_{s}^{-}$ open charm by the statistical uncertainties of their related cross sections. It is found that the uncertainty  arising from this source is 0.1\% which is small enough to be neglected.
	\item $\pi^{+}$ tracking/PID efficiency. The tracking efficiency for $\pi^{+}$ mesons  is studied with a $e^+e^-\to K^+ K^- \pi^+ \pi^-$ control sample. The data-MC tracking efficiency ratios for $\pi^+$ from $D_s^+$ is $1.000\pm0.003$ and that for $\pi^+$ ($\pi^-$) from $\eta^{\prime}$ are $0.988\pm0.008$ ($0.983\pm0.008$). The PID efficiency for $\pi^{+}$ mesons is studied with $e^+e^-\to K^+ K^- \pi^+ \pi^- (\pi^0)$ and $e^+e^-\to \pi^+ \pi^- \pi^+ \pi^-(\pi^0)$ control samples. The data-MC PID efficiency ratios for $\pi^+$ from $D_s^+$ and $\pi^{\pm}$ from $\eta^{\prime}$ are $0.996\pm0.003$ and $0.992\pm0.002$, respectively. Thus, the systematic uncertainties associated with the total charged-particle tracking (PID) efficiency is determined to be 1.9\% (0.7\%).
	\item $\pi^{0}$, $\eta$ reconstruction. The systematic uncertainty associated with the $\pi^{0}$ reconstruction efficiency is investigated by using a control sample of the process $e^+e^-\to K^+K^-\pi^+\pi^-\pi^0$. The same selection criteria described in Sec.~\ref{ST-selection} are used to reconstruct the two kaons and the two pions. The recoiling mass distribution of $K^+K^-\pi^+\pi^-$ is fitted to obtain the total number of $\pi^0$s and the $\pi^0$ selection is applied to determine the number of reconstructed $\pi^0$ mesons. The average ratio between data and MC efficiencies of $\pi^0$ reconstruction, weighted by the corresponding momentum spectra, is estimated to be $1.006 \pm 0.009$. Similarly, the average ratio between data and MC efficiencies of $\eta$ reconstruction is estimated to be $1.011 \pm 0.010$. After correcting the efficiencies, the systematic uncertainties associated with reconstruction efficiencies are 0.9\% for $\pi^0$ and 1.0\% for $\eta$ mesons. 
	\item MC sample size. The uncertainty arising from the finite MC sample size is obtained by $\sqrt{\begin{matrix} \sum_{\alpha} (f_{\alpha}\frac{\delta_{\epsilon_{\alpha}}}{\epsilon_{\alpha}}\end{matrix}})^2$, where $f_{\alpha}$ is the tag-yield fraction, and $\epsilon_{\alpha}$ and $\delta_{\epsilon_{\alpha}}$ are the signal efficiency and the corresponding uncertainty of tag mode $\alpha$, respectively.
	\item Amplitude model. The uncertainty from the amplitude model is estimated by varying the amplitude-model parameters. For the mass and width of $\rho^+$ resonance, we sample them with a Gaussian distribution in which the mean and width are set to the corresponding known value and uncertainty from PDG~\cite{Zyla:2020zbs}. Meanwhile, we uniformly vary the effective radii of Blatt-Weisskopf Barrier within the range $[2.0, 4.0]$ GeV/$c^{-1}$ for intermediate resonances and $[4.0, 6.0]$ GeV/$c^{-1}$ for $D_s$ mesons. The distribution of 600 efficiencies resulting from this variation is fitted by a Gaussian function and the fitted width divided by the mean value is taken as an uncertainty.
	\item Peaking background. 
	The uncertainties caused by peaking background is studied by varying the BF of $D^+_s \to \pi^+ \eta \omega_{\pi^{+}\pi^{-}\pi^{0}}$ from $0.85\%$ to $1.39\%$ based on the precision of the measured branching ratio~\cite{Zyla:2020zbs}. The shift in DT yield is 0.2\%, which is taken as the corresponding uncertainty.
\end{itemize}

All of the systematic uncertainties are summarized in Table~\ref{tab:bfsys}.
Adding them in quadrature gives a total systematic uncertainty in the BF
measurement of 2.9\%.
\begin{table}[htbp]
	\caption{Systematic uncertainties in the BF measurement.}
	\centering
	\begin{tabular}{cc}
		\hline
		Source   &Uncertainty(\%) \\
		\hline
		The number of ST $D^{+}_s$                  &0.9\\
		Tracking                                    &1.9\\         
		PID                                         &0.7\\         
		$\pi^0$ reconstruction                      &0.9\\
		$\eta$ reconstruction                       &1.0\\
		MC statistics                               &0.2\\
		Amplitude model                             &0.4\\
		Peaking background                          &0.2\\
	    BF of $\eta^{\prime} \to \pi^+\pi^-\eta$          &1.2\\
		BF of $\eta \to \gamma \gamma$              &0.5\\
		\hline
		Total                                       &2.9\\
		\hline
	\end{tabular}
	\label{tab:bfsys}
\end{table}

\section{Summary}
This paper presents the amplitude analysis of the decay $D^+_s\to\pi^+\pi^0\eta^{\prime}$ with 6.32 fb$^{-1}$ of $e^+e^-$ collision data samples at $\sqrt{s} = 4.178-4.226$ GeV. The mode $D^+_s \to \rho\eta^{\prime}$ is found to be the main intermediate process contributing to this final state. In addition, we also report the upper limits of the BFs of $S-$wave and $P-$wave nonresonant components of $D^+_s\to \pi^+\pi^0\eta^\prime$ to be $\mathcal{B}(D^+_s\to (\pi^+\pi^0)_{S}\eta^{\prime}) < 0.10\%$ and $\mathcal{B}(D^+_s\to (\pi^+\pi^0)_{P}\eta^{\prime}) < 0.74\%$ at the 90\% confidence level, respectively.

We also measure $\mathcal{B}(D^+_s\to\pi^+\pi^0\eta^{\prime})=(6.15\pm0.25\pm0.18)\%$ which is consistent within 1$\sigma$ of the CLEO result $\mathcal{B}(D^+_s\to\pi^+\pi^0\eta^{\prime})=(5.6\pm0.5\pm0.6)\%$ but has a significantly improved precision. Furthermore, the branching fraction of the $D^+_s \to\rho^+ \eta^{\prime}$ decay is $(6.15\pm0.25(\rm stat.)\pm0.18(\rm syst.))\%$ based on the amplitude analysis results. This result is more than $3\sigma$ above current theoretical predictions and suggests that other contributions, such as, QCD flavor-singlet hairpin amplitude~\cite{Cheng:2011qh},  should be taken into account.

\acknowledgments
The BESIII collaboration thanks the staff of BEPCII and the IHEP computing center for their strong support. This work is supported in part by National Key Research and Development Program of China under Contracts Nos. 2020YFA0406400, 2020YFA0406300; National Natural Science Foundation of China (NSFC) under Contracts Nos. 11625523, 11635010, 11735014, 11775027, 11822506, 11835012, 11875054, 11935015, 11935016, 11935018, 11961141012, 12192260, 12192261, 12192262, 12192263, 12192264, 12192265; the Chinese Academy of Sciences (CAS) Large-Scale Scientific Facility Program; Joint Large-Scale Scientific Facility Funds of the NSFC and CAS under Contracts Nos. U1832204, U1732263, U1832207, U2032104; CAS Key Research Program of Frontier Sciences under Contracts Nos. QYZDJ-SSW-SLH003, QYZDJ-SSW-SLH040; 100 Talents Program of CAS; INPAC and Shanghai Key Laboratory for Particle Physics and Cosmology; ERC under Contract No. 758462; European Union Horizon 2020 research and innovation programme under Contract No. Marie Sklodowska-Curie grant agreement No 894790; German Research Foundation DFG under Contracts Nos. 443159800, Collaborative Research Center CRC 1044, FOR 2359, FOR 2359, GRK 214; Istituto Nazionale di Fisica Nucleare, Italy; Ministry of Development of Turkey under Contract No. DPT2006K-120470; National Science and Technology fund; Olle Engkvist Foundation under Contract No. 200-0605; STFC (United Kingdom); The Knut and Alice Wallenberg Foundation (Sweden) under Contract No. 2016.0157; The Royal Society, UK under Contracts Nos. DH140054, DH160214; The Swedish Research Council; U. S. Department of Energy under Contracts Nos. DE-FG02-05ER41374, DE-SC-0012069.

\bibliographystyle{JHEP}
\bibliography{references}
\clearpage
\collaboration{BESIII Collaboration}
\large
The BESIII Collaboration\\
\normalsize
\\M.~Ablikim$^{1}$, M.~N.~Achasov$^{10,b}$, P.~Adlarson$^{67}$, S. ~Ahmed$^{15}$, M.~Albrecht$^{4}$, R.~Aliberti$^{28}$, A.~Amoroso$^{66A,66C}$, M.~R.~An$^{32}$, Q.~An$^{63,49}$, X.~H.~Bai$^{57}$
% ...
\\
\vspace{0.2cm} {\it
$^{1}$ Institute of High Energy Physics, Beijing 100049, People's Republic of China\\
% ...
$^{4}$ Bochum Ruhr-University, D-44780 Bochum, Germany\\
% ...
$^{10}$ G.I. Budker Institute of Nuclear Physics SB RAS (BINP), Novosibirsk 630090, Russia\\
% ...
$^{15}$ Helmholtz Institute Mainz, Staudinger Weg 18, D-55099 Mainz, Germany\\
% ...
$^{28}$ Johannes Gutenberg University of Mainz, Johann-Joachim-Becher-Weg 45, D-55099 Mainz, Germany\\
% ...
$^{32}$ Liaoning Normal University, Dalian 116029, People's Republic of China\\
% ...
$^{49}$ State Key Laboratory of Particle Detection and Electronics, Beijing 100049, Hefei 230026, People's Republic of China\\
% ...
$^{57}$ University of Jinan, Jinan 250022, People's Republic of China\\
% ...
$^{63}$ University of Science and Technology of China, Hefei 230026, People's Republic of China\\
% ...
$^{66}$ University of Turin and INFN, (A)University of Turin, I-10125, Turin, Italy; (B)University of Eastern Piedmont, I-15121, Alessandria, Italy; (C)INFN, I-10125, Turin, Italy\\
$^{67}$ Uppsala University, Box 516, SE-75120 Uppsala, Sweden\\
% ...
$^{b}$ Also at the Novosibirsk State University, Novosibirsk, 630090, Russia\\
% ...
}

\end{document}